\documentclass[letterpaper,twocolumn,10pt]{article}
\usepackage{usenix}
\usepackage{tikz}
\usepackage{amsmath}
\usepackage{booktabs}
\usepackage{caption}

\usepackage[T1]{fontenc}
\usepackage[scaled=0.75]{beramono}
\usepackage{listings}
\usepackage{textcomp} 
\usepackage{placeins}
\usepackage{algorithm}
\usepackage[noend]{algpseudocode}
\usepackage{multirow}
\usepackage{url}
\usepackage{array}
\usepackage{tabularx}

\newcommand{\circnum}[1]{%
  \tikz[baseline=(char.base)]{
    \node[shape=circle, fill=black, text=white, inner sep=0.4pt, minimum size=0.8ex] (char) 
      {\scriptsize #1};
  }%
}

\newcommand{\oreq}{\mathrel{|=}}

\newcommand{\toolInstr}{ZIMPAF}
\newcommand{\toolFuzz}{RedPhuzz} 

\newcommand{\CC}[1]{\emph{\textbf{{\color{blue} [CC says: #1]}}}}

\begin{document}

\date{}

\title{\toolInstr{} \& \toolFuzz{}: High-fidelity Web Application Fuzzing via Branch, Language Construct, and Function Call Monitoring}


\author{
{\rm Tennov Simanjuntak}\\
University of Texas at Arlington
\and
{\rm Christoph Csallner}\\
University of Texas at Arlington
}


\maketitle

\begin{abstract}
We present \toolInstr{}, runtime interpreter instrumentation, and \toolFuzz{}, a fuzzer, to realize high-fidelity fuzzing. \toolInstr{} implements multi-granular interceptors to collect traces of branch coverage, errors and exceptions, functions and language constructs, and input-tainted branches. The framework identifies tainted vulnerable functions, marking them as high-value fuzzing targets, and  untainted vulnerable functions using in-situ static analysis called \emph{backward constant probe}, allowing them to be skipped, without pre-fuzzing static taint analysis. The traces are utilized by \toolFuzz{} to perform targeted function- and input-level fuzzing that goes beyond traditional coverage-based fuzzing supported by three novel mutation strategies: sanitization-aware, input-in-branch-aware, and data-type-aware mutation. Furthermore, we introduce multi-stage vulnerability detection that extends beyond error- and exception-based detection, capable of detecting silent vulnerabilities. We evaluate \toolFuzz{}'s performance against its predecessor, Phuzz, on 86~test cases across six benchmark web applications. \toolFuzz{} detects all vulnerabilities, whereas Phuzz fails to detect~16. \toolFuzz{} is 77\% faster than Phuzz despite performing more tasks. \toolInstr{} is faster than Phuzz's instrumentation (PCOV and UOPZ) while writing significantly more data. \toolInstr{} achieves 2.1 to 41.22 times higher throughput than PCOV and UOPZ across five sampled benchmarks, and 0.87$\times$ for one.      
\end{abstract}

\section{Introduction}
\label{sec:intro}

The central challenge for grey-box web application fuzzing is to maximize the amount and quality of execution feedback from application internals without introducing excessive overhead. Despite much progress, state-of-the-art fuzzers continue to exhibit several weaknesses rooted in their failure to reflect the key difference between native and web applications, resulting in limited execution-context awareness and inefficient instrumentation. Furthermore, they lack essential web-domain knowledge. Unlike native applications, server-side web applications, such as PHP, run on an interpreter that preserves high-level semantics in compile-time and runtime data structures~\cite{appel1998}, which can be leveraged to provide high-fidelity runtime semantics feedback to guide fuzzing. The interpreter is eventually executed by the CPU, introducing additional execution overhead compared with native applications; thus efficient instrumentation is essential.  



Web applications are the backbone of customer engagement and business processes, making them a common entry point for security attacks. Verizon's 2025 DBIR ranks basic web application attacks third among incident patterns~\cite{verizon2026dbir}. Although web vulnerabilities can emerge at different levels, many manifest through unsafe or incorrect function-level implementations~\cite{owasp2025}.

\phantomsection
\label{sec:lack-of-context}
In web applications, vulnerabilities often emerge from functions or function-like syntax, such as PHP language constructs~\cite{phpnet}. Therefore, effective fuzzing should operate at the function level, specifically targeting vulnerable functions along the execution path, with mutations deterministically guided by their input-tainted parameters. This requires instrumentation to expose function-level execution semantics to fuzzers. The fuzzer must also skip invulnerable functions whose parameters untainted by user input and that are safely executed as part of the call-sequence semantics such as \texttt{prepare}, \texttt{bind}, and \texttt{execute} for parameterized queries. Furthermore, web application code commonly calls input transformation functions, such as sanitizations, decoding, and string replacement and matching prior to vulnerable function, requiring the mutation to satisfy the transformation to reach the target function~\cite{stuttard2011}. State-of-the-art fuzzers lack awareness of function-level execution-context~\cite{vanRooij2021,zhao2022,trickel2023, wang2025} and rely primarily on code coverage, errors, and exceptions, limiting their mutation strategies to reach vulnerable functions. Nevertheless, Cefuzz~\cite{zhao2022} and Predator~\cite{wang2025} employ heavyweight static taint analysis and block-distance analysis for directed fuzzing to provide feedback on the execution distance to vulnerable functions. However, it likely fails with common web-application input transformations, as the fuzzer cannot satisfy the conditions needed to reach target functions without transformation awareness. It is also difficult to scale in production environments with frequent application updates and testing. Phuzz~\cite{neef2024} and Atropos~\cite{guler2024} augment feedback by monitoring calls to PHP userland vulnerable functions, a widely used technique in malware analysis~\cite{egele2012,ormeir2019}. However, feedback signals only verifies that errors and exceptions are triggered by user input and do not guide mutation. They also succumbs to vulnerable functions guarded by input transformation functions and miss vulnerabilities emerging from language constructs (such as \texttt{include} or \texttt{eval}), which are directly implemented as opcodes instead of function calls.

Furthermore, reliance on errors and exceptions as the sole indicator of vulnerabilities is limited because web application code can allow only well-formed input to reach vulnerable functions without triggering errors, leaving flaws undetected. Function-level execution semantics helps fuzzer overcome this by inspecting whether submission of malicious well-formed input results in a successful execution, implying the vulnerability

\phantomsection
\label{pg:web_domain_knowledge}
Web applications are often supported by databases as backend storage. An effective fuzzer should utilize database-contextual semantics to perform  mutation when the target are database functions. It can generate type-conformed input to reach vulnerable functions and deliberately violate type constraints to trigger bugs caused by database constraint violations. State-of-the-art fuzzers lack this capability.  

\phantomsection
\label{pg:inefficient-instrumentation}
As the enabler of grey-box fuzzing, efficient instrumentation is essential. However, state-of-the-art fuzzers' instrumentation is inefficient. They employ either userland instrumentation~\cite{vanRooij2021,zhao2022}, or interpreter instrumentation by patching the interpreter~\cite{trickel2023} or using PCOV or Xdebug~\cite{krakjoepcov, rethans} extension~\cite{guler2024,neef2024}. Userland instrumentation requires heavyweight static analysis to insert instrumentation code, which increases code size and (due to frequent code updates) does not scale in production. Likewise, their interpreter instrumentation sustains substantial overhead by globally intercepting the interpreter’s main execution loop, thereby hooking every userland opcode, potentially hundreds of thousands or millions per run. This overhead is amplified by fuzzing's large number of iterations. Predator, built on top of Witcher, claims to use selective instrumentation to achieve both effectiveness and efficiency~\cite{wang2025}; however, our analysis reveals that instrumentation hooking remains global, with selectivity applied only in the trace handler. Phuzz's use of UOPZ~\cite{krakjoeuopz} to monitor userland function calls is also inefficient due to per-request hook installation, repeated closure creation and execution, and function dispatch interception via opcode-handler overloading that effectively intercepts all userland function calls.

To address the weaknesses of state-of-the-art fuzzers, we propose a novel, efficient, and effective grey-box web application fuzzing framework consisting of \toolInstr{}, a runtime interpreter instrumentation, and \toolFuzz{}, a redesigned fuzzer that is execution-context-aware and incorporates essential web-domain knowledge. The framework goes beyond code-coverage-guided and error- and exception-based fuzzing by leveraging the rich compile-time and runtime semantics exposed by the PHP interpeter (i.e., the Zend engine~\cite{phpint}). \toolInstr{} employs pure dynamic analysis based on multi-granular function-call and opcode-execution monitoring to provide high-fidelity runtime semantic feedback to guide input mutation and to detect vulnerabilities and bugs. Importantly, the framework deliberately avoids static taint analysis prior to fuzzing, such as~\cite{zhao2022,wang2025} which is inefficient, particularly in a production environment where frequent web-application updates require repeated analysis of the evolving codebase. To achieve this, \toolInstr{} employs a novel in-situ static analysis called \emph{backward constant probe}, over the Zend's compile-time data structure to identify functions and language constructs untainted by user input while hooks still active. Furthermore, it performs runtime tracking to identify branches influenced by user input by comparing branch operands with elements of Zend's data structures that store user inputs. Lastly, \toolFuzz{} identifies input-tainted sinks by using string matching heuristics. 
Following are this paper's main contributions.
\begin{itemize}
  \item A novel multi-granular instrumentation to collect traces of branch coverage, errors and exceptions, vulnerable functions and language constructs, sanitization functions, and input-tainted branches as feedback for fuzzers. The instrumentation is deployed as a PHP extension. 
  \item A novel in-situ static analysis called \emph{backward constant probe} to identify invulnerable functions and language constructs. 
  \item A novel, highly-targeted function-level fuzzing that enables deterministic mutation when function-level semantics are available.  
  \item Three novel mutation techniques: sanitization-aware, input-tainted-branch-aware, and data-type-aware mutation.
  \item A novel, multi-stage vulnerability detection based on errors and exceptions, function-execution semantics, and safe-sequence semantics. 
  \item Empirical comparison with the state-of-the-art (its predecessor: Phuzz) on six benchmark web applications comprising 86 test cases: Phuzz misses 16, our framework detects all vulnerabilities. Our framework is also 77\% faster than Phuzz despite performing additional tasks and producing more instrumentation data.
\end{itemize}
To foster further research in this field, we open-source the framework implementation at this link: \url{https://github.com/tennovs/zimpaf_redphuzz}. 

\section{Motivating Examples}
\label{sec:motivating_examples}

To highlight the importance of the execution-context awareness and web domain knowledge, we present two motivating examples given in Listings~\ref{lst:listing-1} and~\ref{lst:listing-2}. 

\begin{lstlisting}[language=PHP,
				   caption={Invulnerable and vulnerable functions with decoding and sanitization functions},
                   captionpos=b, 
                   label={lst:listing-1},
                   numbers=none,
                   breaklines=true,
                   breakatwhitespace=false]
<?php
define("DB_ROOT", "mysql_db/");
include (DB_ROOT . 'db.php'); //not vuln (*@\circnum{1}@*)
$name = base64_decode($_GET['name']); (*@\circnum{2}@*)
$file = base64_decode($_GET['file']); (*@\circnum{3}@*)
if(!is_numeric($name) && !is_numeric($file)){ (*@\circnum{4}@*)
  if(fnmatch("schedule*", $file)){	(*@\circnum{5}@*)
    $query  = "SELECT role from users where username = '$name'"; //tainted (*@\circnum{6}@*)
    $result = mysqli_query($conn, $query); //vuln (*@\circnum{7}@*)
    $row    = mysqli_fetch_assoc($result);
    $role   = $row['role'];
    $path   = $role . "/" . $file; //tainted (*@\circnum{8}@*)
    include("$path"); //vuln (*@\circnum{9}@*)
  }else{
    echo "ERROR: No file name started with $file";
    exit;
  }
  $query  = "SELECT username FROM users WHERE role = 'branch_manager'"; 
  $result = mysqli_query($conn, $query); //not vuln (*@\circnum{10}@*)
  $row = mysqli_fetch_assoc($result);
  $manager = $row['username'];
  echo $manager . "<br>";
  echo "Schedule of $role is verified by $manager.";
}
?>
\end{lstlisting}

In Listing~\ref{lst:listing-1}, calls at \circnum{1} and \circnum{10} are not vulnerable because their parameters originate from constant values; thus, no user input can reach them. The parameter of \texttt{include} construct at \circnum{1} is a concatenation of a constant defined in the previous \texttt{define} statement and another constant within itself. At \circnum{10}, \texttt{mysqli\_query} accepts a \texttt{query} parameter from a literal string in the preceding statement. A fuzzer that monitors function call, but incapable of identifying parameter sources may mistakenly treat inputs with these constructs and function as interesting, leading to unnecessary fuzzing and wasted computation. Therefore, it should be able to identify invulnerable constructs and functions and remove them from the fuzzing list. In contrast, functions at \circnum{7} and \circnum{9} are reachable from user requests, tainted at \circnum{6} and \circnum{8}, and vulnerable, since no sanitization is performed for their parameters. However, the fuzzer needs to provide the correct inputs to pass three blockages for the execution to reach them. Each input parameter must be base64-encoded so that statements at \circnum{2} and \circnum{3} can decode them correctly. Both parameters must be non-numeric (\circnum{4}), and finally, the \texttt{file} parameter must begin with \texttt{schedule} (\circnum{5}). 

State-of-the-art-fuzzers~\cite{vanRooij2021,zhao2022,trickel2023,guler2024, neef2024,wang2025} miss all or part of this knowledge making them ineffective when mutating the inputs. Atropos~\cite{guler2024}, despite being able to identify string comparison to solve the conditional statement such as that with \texttt{fnmatch}, it cannot effectively generates input to satisfy a series of sanitization given in Listing~\ref{lst:listing-1}. A smarter fuzzer must be able to leverage this information to perform effective mutation tailored to the tainted parameters. It can also parse \texttt{query} parameter to perform even more targeted mutation and determine the exploitabilty of \texttt{mysqli\_query} at \circnum{7}. By parsing the query, the fuzzer knows the parameter is quoted, allowing it to submit only quoted parameter, skipping numeric parameter to trigger a vulnerability. Further, it can signal whether the vulnerability is actually exploitable by checking if the malicious input parameter eventually turns into logic and the function's return value is a success. Likewise, it can gather database metadata to identify data types to perform type-conformed mutation and to probe bugs by violating the data‑type or domain‑constraint violations. While Atropos and Phuzz can verify whether input parameters reach vulnerable functions when errors or exceptions occur, they lack these capabilities.

Equally important, a fuzzer monitoring function calls must recognize that potentially vulnerable functions or methods are indeed safe because they are executed in a sequence that prohibits maliciously-crafted inputs to trigger vulnerabilities, allowing their fuzzing to stop. Listing~\ref{lst:listing-2} illustrates this with two examples. The \texttt{prepare} method at \circnum{1} is executed in a safe sequence: the \texttt{query} parameter originates from a literal string--query with a placeholder \texttt{?}--in the previous statement; \texttt{bind\_param} binds the \texttt{\$odate} variable to the prepared statement; and finally, \texttt{execute} executes the prepared statement. This call-sequence semantic ensures that the query parser never turns user input into logic that might trigger an SQL injection vulnerability. Nevertheless, it may have bugs, such as a domain violation caused by a long string.

\texttt{mysqli\_query} function at \circnum{2} is also secure because the user input \texttt{\$\_GET['prodname']} is sanitized using \texttt{mysqli\_real\_escape\allowbreak\_string} and inserted into \texttt{\$query} as a quoted string, preventing it from being interpreted as SQL logic. These examples illustrate the importance of modeling web-domain knowledge in a fuzzer to skip safe functions where fuzzing is meaningless and focus on high-value functions likely to trigger vulnerabilities or bugs.

In contrast, while \texttt{mysqli\_query} at \circnum{4} is free from XSS injection, it is vulnerable to SQL injection. \texttt{htmlspecialchars} with \texttt{ENT\_QUOTES} and \texttt{UTF-8} parameter guards the function from an XSS injection by transforming user input into characters that are not executable as a script by the user's browser, while no sanitization protects SQL query. However, to trigger SQL vulnerability, the execution must satisfy the conditional statement with a "\textgreater" operator at \circnum{3} whose operand is user input. Fuzzers unware of this branch are likely to fail because their mutation cannot effectively generate satisfying input to reach \texttt{mysqli\_query}. 

\vspace{1.0em}
\begin{minipage}{\columnwidth}  
\begin{lstlisting}[language=PHP,
  caption={Safely executed and vulnerable userland functions (code excerpt).}, 
  captionpos=b,
  numbers=none,
  label={lst:listing-2}]   
if(isset($_GET['odate'])){
  $odate = $_GET['odate'];
  $query = "SELECT o.order_id, o.product_name, o.order_date, u.username FROM orders o JOIN users u ON o.user_id = u.user_id WHERE DATE(o.order_date) = ?";
  $stmt = $conn->prepare($query); //safe, but has bug  (*@\circnum{1}@*)
  $stmt->bind_param("s", $odate);
  $stmt->execute();
  $result = $stmt->get_result();
  /* ... */
}
if(isset($_GET['prodname'])){
  $prodname = mysqli_real_escape_string($conn,$_GET['prodname']);
  $query = "SELECT * FROM orders WHERE product_name = '$prodname'";
  $result = mysqli_query($conn, $query); //safe (*@\circnum{2}@*)
  /* ... */
}
if(isset($_POST['add_user'])){
  $username = $_POST['add_user'];
  $income = $_POST['income'];
  $age = $_POST['age'];
  if ($_POST['age'] > 17){	(*@\circnum{3}@*)
    $username = htmlspecialchars($username, ENT_QUOTES, 'UTF-8'); 
    $query = "INSERT INTO users (username, age, income) VALUES ('$username', $age, $income)";
    $result = mysqli_query($conn, $query); //vulnerable (*@\circnum{4}@*)
  }
}
\end{lstlisting}
\end{minipage}

\section{Design}
\label{sec:design}

\begin{figure*}[h!t]
    \centering
    \includegraphics[width=\textwidth, trim={.2in .1in .2in .1in}, clip]{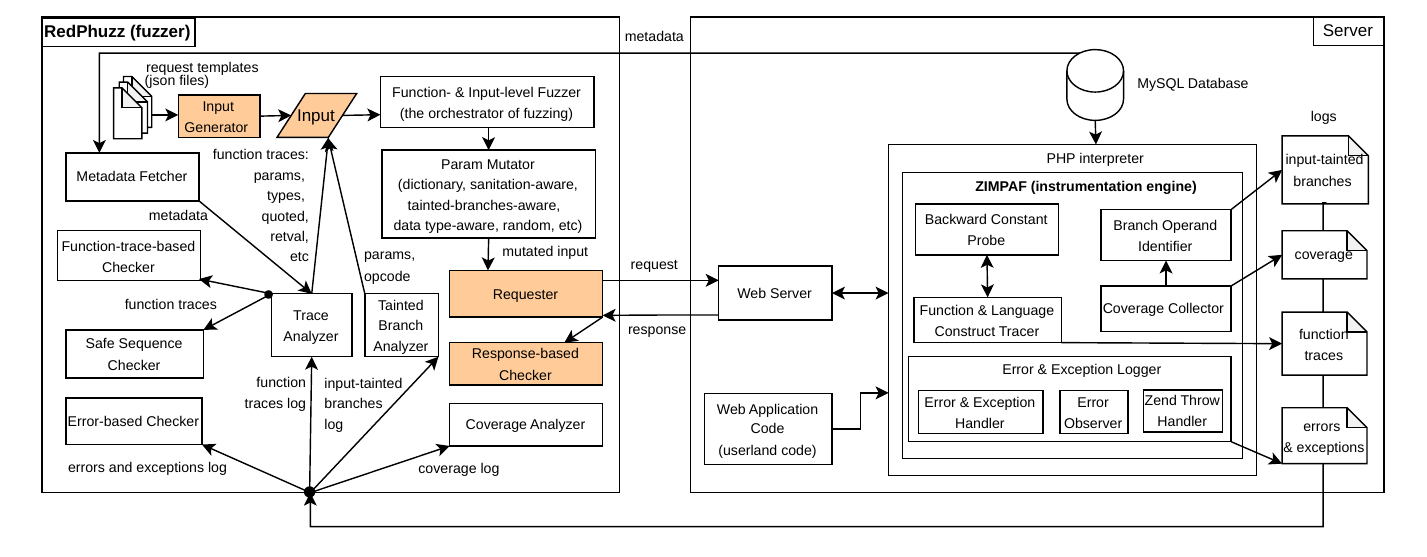}
    \caption{System architecture: Front-end fuzzer (left) with Phuzz~\cite{neef2024} components shaded and back-end instrumentation (right).  
    } 
    \label{fig:architecture}
\end{figure*}

Figure~\ref{fig:architecture} shows the system architecture of our fuzzing framework to address the limitation of state-of-the-art fuzzers. The framework consists of RedPhuzz, the front-end fuzzer and ZIMPAF, the back-end instrumentation, connected by a web server. The architecture incorporates the database to provide the fuzzer with contextual-information during the execution of userland database functions. We target PHP web applications because of the architectural transparency, modularity, and programmability of its interpreter (the Zend Engine~\cite{phpint}) and as PHP remains by far the most-widely used server-side technology for web applications~\cite{w3techs}.

\subsection{Back-end Instrumentation: \toolInstr{}}

To achieve efficient and effective runtime interpreter instrumentation, we leverage the Zend Interpreter’s multi-granular interception modules and its rich compile-time and runtime data structures. The instrumentation consists of three interceptors: a coverage collector, an error and exception logger, and a function and language construct tracer. The first and third employ branch operand identifier and backward constant probe, respectively. 
\phantomsection
\label{ssec:architecture}

\subsubsection{Coverage Collector \& Branch Operands Identifier}

We opt for efficient, fine-grained opcode-execution monitoring~\cite{phpint} to collect branch coverage of userland code. This design eliminates the unnecessary overhead incurred by existing state-of-the-art fuzzers~\cite{trickel2023,guler2024,neef2024,wang2025}, which intercept all opcodes in the interpreter's main execution loop. Instead, \toolInstr{} hooks branch opcode handlers, where execution may diverge into multiple paths; thus, excluding opcodes that do not create distinct execution paths. \toolInstr{} overwrites Zend's branch-opcode handlers with pointer to its own handlers once during extension loading, avoiding per-request setup overhead incurred by PCOV~\cite{krakjoepcov} and Xdebug~\cite{rethans} extension. This design also eliminates code patching required by~\cite{trickel2023,guler2024,wang2025} to log coverage, which is difficult to maintain across interpreter versions.

\phantomsection
\label{coverage-triples}

An \emph{execution path} is represented as a sequence of \emph{(file path, branch-opcode userland line number, branch-opcode outcome)} triples. As single userland conditional statement may translate into multiple branch opcodes, multiple triples may represent a single conditional. We use this representation instead of fully evaluating those triples to reduce computational overhead, as execution paths provide inexpensive yet robust signals for distinguishing new inputs from previously explored ones.

Further, coverage collector utilizes the branch operand identifier to identify branches influenced by user input, which is crucial to satisfy conditionals to reach vulnerable functions, such as the ones at \circnum{3} and \circnum{4} in Listing~\ref{lst:listing-2}. The branch operand identifier matches branch operand pointers against inputs stored in the PHP superglobal array. It falls back to value matching when pointer matching fails, such as when \texttt{ZEND\_JMP*} assigns an input parameter to a temporary variable for optimization to avoid updating the reference count.

\subsubsection{Error and Exception Logger}

The Zend interpreter behavior in propagating errors and exceptions to userland code varies. It depends on the existence of userland error function or whether the Zend interpreter performs a bailout--bypassing the normal execution flow--to return to a safe state immediately during errors or exceptions~\cite{phpint}. Similarly, the interpreter may either write errors and exceptions to response or suppress them entirely. A robust errors and exceptions logger must account for this behavior. For this purpose, we opt for decomposing the logger into three components and avoid inferring errors and exceptions from responses as it is inefficient and can fail too. The first component is the error and exception handler, two separate handlers that intercept the Zend's main mechanism for propagating errors (\texttt{zend\_error\_cb} and \texttt{zend\_throw\_exception}). The second is an error observer function, registered to the error observer chain, which is useful when \texttt{zend\_error\_cb} is not called, such as during non-fatal error like warning and notice. The third component intercepts \texttt{ZEND\_THROW} opcode handler, which is always called to initiate the creation of throwable object, the last resort if the interpreter decides to bailout. 

Our strategy is more robust than patching the error dispatcher, \texttt{php\_\allowbreak error\_cb} or its companion, \texttt{php\_verror} as required by Atropos~\cite{guler2024}), which can fail to capture exceptions. Similarly, Phuzz \cite{neef2024}, which relies on UOPZ extension~\cite{krakjoeuopz} also misses compilation errors, suppressed warnings, and exceptions caught by application \texttt{try-catch} blocks.

\subsubsection{Function and Language Construct Tracer and Backward Constant Probe}
\label{sec:fung-lang-tracer}

Function and language construct tracer monitor the execution of userland functions--vulnerable and sanitization functions--and language constructs, respectively. Function tracer hooks the handlers of internal functions that execute userland functions inside the Zend interpreter, while language construct tracer hooks the construct opcodes~\cite{phpint}. This fine-grained design eliminates the code patching required by Atropos~\cite{guler2024} and avoids the overhead of global hooking associated with UOPZ~\cite{krakjoeuopz} (Phuzz’s instrumentation), which hooks function dispatch opcodes. More importantly, the design overcomes both inability to instrument language constructs. The tracing mechanism is customized per vulnerability class to track functions capable of triggering specific security flaws.

The trace log contains parameters, return values, location (file and line number), and other attributes which are essential to guide mutation and uncover vulnerabilities and bugs. Function or language-construct parameters that may receive user inputs are sinks; a sink is tainted when it receives user input and untainted otherwise.

\phantomsection
\label{sec:originate}

The tracer employs in-situ static analysis, the \emph{backward constant probe} to mark vulnerable functions and language constructs whose sink originates from constant values, indicating that they cannot be tainted. This marking allows \toolFuzz{} to skip them during function-level fuzzing, such as \texttt{include} at \circnum{1} and \texttt{mysqli\_query} at \circnum{10} in Listing~\ref{lst:listing-1}. \emph{As all vulnerable sinks are strings, we define that a sink originates from constant values when it is either a single hard-coded string literal or a string literal constructed from string operations that may involve non-string values}. With this mark added to the trace log, the fuzzer filters out invulnerable functions. The marking is also useful to check if a vulnerable function is executed in a safe sequence, e.g., Listing~\ref{lst:listing-2} \texttt{prepare} at \circnum{1}.

The backward constant probe is efficient because it relies on a few string transformation opcodes stored in the Zend opcode array to track the origin of the string in the sink. It also bypasses sink identification because the probe starts while the function and language construct's hook is active. This mechanism avoids performing static taint analysis prior to fuzzing as required by~\cite{zhao2022,wang2025}. To our knowledge, this is the first backward constant probe applied to web-application fuzzing and potentially vulnerability discovery more broadly.

We leverage the tracer that monitors functions vulnerable to SQL injection and code execution to log errors for two reasons. First, the Zend error and exception handler does not produce a log containing an error number capable of distinguishing SQL injection from bugs. The tracer thus retrieves the error number and related metadata directly from the database driver. An error number indicating parsing error signals SQL injection. Fuzzers lacking this capability will produce more SQL injection false positives. Second, the Zend interpreter suppresses errors generated by code execution functions from its main error handler, writing them to the response instead. Thus, \toolInstr{} logs the functions whose return values can indicate an error. For functions whose return values cannot be used to determine an error, \toolInstr{} safely simulates their execution and retrieves the status from a resource stream. This avoids external execution required by Witcher~\cite{trickel2023}, which executes commands inside \texttt{dash}, a Debian shell.

The coverage, input-tainted branches, error and exception log, and function and language construct traces are written to shared files accessible for the fuzzer.

\subsection{Front-end Fuzzer: \toolFuzz{}} 

\toolFuzz{} is a redesigned version of Phuzz~\cite{neef2024} consisting of 13 components, excluding request templates, a collection of JSON files representing valid HTTP requests. It inherits four components as shown by shaded components in Figure~\ref{fig:architecture}. The fuzzing starts when the input generator parses a request template to create an input object. The function- or input-level fuzzer then mutates input parameters in the object and calls the requester. The requester creates an HTTP request and submit it to the server where \toolInstr{} performs instrumentation and writes the logs to be consumed by \toolFuzz{}. 

\subsubsection{Coverage Analyzer}   

The coverage analyzer loads and hashes the current input's path coverage log. The fuzzer uses the \textbf{path hash} to identify a new path by (constant time) set membership checking in the set of previous path hashes. If the path hash is not unique (path previously explored), the analyzer returns immediately. If the hash is unique (new execution path), the analyzer computes the path's unique triples and concatenates it with the aggregated triples to reflect the existing total coverage. Lastly, it creates a new input for the new path. The hash check augments the coverage as cheap signals to identify new inputs, particularly since vast majority of mutated inputs result in identical execution paths during fuzzing. The design also avoids the unnecessary overhead incurred by Phuzz\cite{neef2024}, which relies on string differences rather than hashes.

\subsubsection{Trace Analyzer and Metadata Fetcher}
\label{sssec:trace-metadata}   

The trace analyzer analyzes the current path's function and language-construct traces, collectively referred to as function traces for brevity, as PHP language constructs behave like function calls. It extracts semantics for vulnerable and sanitization function, such as function name, their parameters and return values, and other essential information such as filenames and line numbers. It classifies potentially vulnerable functions by vulnerability type, including code execution, path traversal, SQL injection, deserialization, and XML external entity. To avoid wasted computation, it excludes functions whose parameters originate from constants, as indicated by the \texttt{sink opline type} attribute set by the instrumentation's backward constant probe. It also detects \texttt{die} and \texttt{exit}, which indicate early script termination.

For vulnerable functions, the trace analyzer then identifies inputs that taint the functions by matching each input against the substring in the function's parameter (sink), marking them as high-value targets, and restricting mutation solely to inputs that taint the sink. Listing~\ref{lst:listing-1} (at \circnum{7} and \circnum{9}) illustrates this, where parameters of vulnerable functions contain substrings supplied by the user via HTTP GET. This approach is similar to Atropos~\cite{guler2024}; both aim to avoid the high cost of pre-fuzzing taint analysis incurred by prior work~\cite{zhao2022, wang2025}. Next, the analyzer reports sanitization applied to inputs and sinks in two forms. First, the input-level report records the sequence of sanitization functions applied to an input. Second, the function-level report records the sequence applied to input-tainted sink before the vulnerable function is called.

Furthermore, for functions vulnerable to SQL injection, the analyzer also identifies the data type of each SQL query parameter tainted by user inputs and whether it is quoted. The fuzzer first parses the SQL query to identify table names, column names, query parameters, and a quoted flag, an indicator whether a query parameter is inside quotes (string) or not (numeric). The analyzer then retrieves the tables' metadata by calling the metadata fetcher and matches the metadata to the parsing result to determine each query parameter's data type and the value of the quoted flag. The results of the trace analyzer are added to the input object.

The sanitization and data type semantics are used by param mutator to perform sanitization-aware and data-type-aware mutation, respectively (see Section~\ref{sssec:param-mutator}).

\subsubsection{Tainted-Branch Analyzer}
\label{sssec:branch-param-analyzer}

The tainted-branch analyzer analyzes the tainted-branch log to extract inputs reaching branch operands and the corresponding opcodes and add them to the input object. Listing~\ref{lst:listing-2} at \circnum{3} illustrates such a branch, where the first operand is a user-supplied parameter passed to the script via HTTP POST. The fuzzer uses this feedback to perform tainted-branch-aware mutation (Section~\ref{sssec:param-mutator}).

\subsubsection{Param Mutator}
\label{sssec:param-mutator}

The first mutation strategy is dictionary-based, applied when function-level semantics are available to inputs that taint the sink. For each vulnerability class, we define a dictionary of malicious malformed and well-formed inputs. For SQL injection, both malformed and well-formed inputs are categorized into quoted (string) and unquoted (numeric) inputs, enabling a more targeted mutation. The malformed dictionary aims to trigger parsing errors or exceptions, which signal a vulnerability. In contrast, malicious well-formed inputs aim to exploit a vulnerable function, when web application code allows only well-formed input to reach the function, rendering no errors or exceptions, while the vulnerability remains.

The second strategy is sanitization-aware mutation, applied when the trace analyzer yields a sanitization report. This strategy takes a pre-sanitization input (e.g., from a dictionary or randomly generated) and reverses the sanitization sequence, starting from the last function and proceeding to the first. Sanitization-aware mutation can satisfy conditions for reaching vulnerable functions or violate them to explore new paths.

The third strategy is tainted-branch-aware mutation, applied when the tainted-branch analyzer identifies input-tainted branches for the current input. This strategy comprises parameter-preserving mutation, which preserves the path condition that allows execution to reach a vulnerable function, and parameter-flipping mutation, which flips the branch to explore new paths.

The fourth strategy is data-type-aware mutation, applied when data-type and quoted-flag semantics for functions vulnerable to SQL injection are available. This strategy performs type-conforming mutation to reach vulnerable functions and type-violating mutation to trigger bugs. 

Besides these strategies, we also define other mutation techniques, including zero-or-empty mutation to generate numeric zeros or empty strings, as well as numeric-only, string-only, and random mutation.

\phantomsection
\label{sssec:mutation-hash}
Unlike Phuzz~\cite{neef2024}, \toolFuzz{} avoids writing every mutation on known paths to disk; instead, it hashes mutations to prevent duplication, saving significant storage.

\subsubsection{Function- and Input-Level Fuzzer}

This is the orchestrator that drives the fuzzing workflow by invoking the parameter mutator, requester, coverage analyzer, and vulnerability checkers in each fuzzing iteration, while lazily invoking the trace analyzer and tainted-branch analyzer only upon discovering new inputs. The function-level fuzzer performs deterministic fuzzing when function-level semantics are available and uses dictionary-based mutation to uncover vulnerabilities and bugs. First, it submits malicious malformed dictionary to trigger parsing errors and exceptions that signal vulnerabilities or bugs. Next, it submits malicious well-formed dictionary to detect silent vulnerabilities when the web application code permits only well-formed inputs to reach vulnerable functions, producing no errors or exceptions. 

When sanitization precedes a vulnerable function, the function-level fuzzer applies sanitization-aware mutation before input submission. Likewise, when vulnerable functions are hidden behind input-tainted branches, it applies tainted-branch-aware mutation before input submission.

We provide a specific function-level fuzzer for functions vulnerable to SQL injection and a generic function-level fuzzer for other vulnerability classes. Because SQL functions performing insert or update operations may serve as sinks for stored XSS, we also define a dedicated fuzzer for stored XSS, which is invoked immediately after SQL-injection fuzzing. Since the trace analyzer may fail to identify sinks in vulnerable functions due to sanitization and tainted-branch, function-level fuzzing is also used to uncover hidden sinks.

The input-level fuzzer is used when function-level semantics are unavailable, stochastically mutating inputs to explore new paths aimed to reach vulnerable functions. Depending on the availability of sanitization and tainted-branch reports, we define three input-level fuzzers: one that performs sanitization-aware mutation prior to input submission, one that performs tainted-branch-aware mutation prior to input submission, and one that performs neither when neither report is available. The input-level fuzzer also uncovers vulnerabilities that do not emerge from function calls, such as reflected XSS, using an XSS-specific malicious dictionary.
 
\subsubsection{Bug and Vulnerability Checker}

\toolFuzz{} employs a multi-stage checker that goes beyond the tradition errors- and exceptions-based detection to discover vulnerabilities and bugs. The checkers are explained as follows.

\noindent\textbf{Error-based Checker:}
The error-based checker detects vulnerabilities and bugs by analyzing the error and exception log and lives in the entire fuzzing iteration. The entries in error and exception log share common attributes: function name, message, and location; the latter is represented by a file path and line number. The identity of errors are represented as error number in error log and by code in exception log. This checker treats an SQL error or exception as a bug, unless it is categorized as a parsing or syntax error (e.g., SQL error no. 1064 in MySQL), which we treat as a vulnerability. State-of-the-art fuzzers~\cite{trickel2023,guler2024,neef2024,wang2025} overlook this, making them prone to many SQL injection false positives. For other vulnerability classes, if the syntax error occurs and originates from a vulnerable function, we treat it as a vulnerability.

\noindent\textbf{Function-trace-based Checker:}
The function-trace-based checker runs after the error-based checker and lives when the malicious well-formed inputs are submitted during dictionary-based mutation. This checker allows \toolFuzz{} to uncover silent vulnerabilities when only well-formed input is allowed to reach the vulnerable function, resulting in no errors being generated. In this case, vulnerability can only be inferred by the function's exploitability. The function-trace-based checker is written for each vulnerability class and uncovers a vulnerability by inspecting whether the return value represents a successful execution and the presence of well-formed input in sink, indicating a vulnerability. Detecting SQL injection vulnerabilities is more difficult than other vulnerability classes as sanitization can tame the malicious well-formed input causing the return value holds a success, which leads the fuzzer to decide vulnerability, while in fact, it is not. To avoid this false positive, the fuzzer employs a heuristic to determine whether the malicious well-formed input eventually turns to SQL logics. If yes, the fuzzer flags the function as vulnerable.  

\noindent\textbf{Safe-sequence Checker:}
The safe-sequence checker verifies the safety of vulnerable function executions, such as those annotated \circnum{1} and \circnum{2} in Listing~\ref{lst:listing-2}. Once a vulnerable function is determined to be safe, the fuzzer ceases all further fuzzing of the function to avoid meaningless computation. To realize this, the checker is equipped with the safe function-call semantics of vulnerable functions executions for each vulnerability class.  \emph{A safe function-call semantic is defined as a sequence of function calls, one which is a vulnerable function, when executed according to the sequence and with certain properties associated to the corresponding functions, ensuring the invulnerability of the vulnerable function}. The safe-sequence checker is also used to verify vulnerabilities detected by the function-trace-based checker to avoid false positive.

\toolFuzz{} hashes new vulnerabilities and bugs and uses the hash to prevent duplication.

\subsection{Algorithms} 




Algorithm~\ref{alg:instrcorepipe} shows the instrumentation core pipeline, which respects both the Zend engine lifecycle and its stack frame. The global module variable \texttt{coverage\_id} identifies and input mutation represented by an HTTP request and serves as log filename. The dynamically allocated \texttt{path\_table} array stores a path recorded by the coverage collector, the \texttt{f\_call\_sequence} array holds the sequence of traces collected by the function and language construct tracer, and the \texttt{tainted\_branch\_log} array stores the details of branches whose operands are input parameters. All of them are written to the filesystem efficiently in batches. 

\texttt{MODULEINIT} installs all the interceptors during module startup to make them alive throughout the Zend lifecycle, ensuring they are installed once for all requests. Each interceptor redirects execution to its respective handler at \circnum{2}, \circnum{3}, \circnum{4}, \circnum{6}, and \circnum{7}. The hooking at \circnum{1} is installed for each monitored function. \texttt{GLOBALSINIT} allocates memory for the \texttt{path\_table}, which is eventually freed by \texttt{GLOBALSHUTDOWN} (not shown). These global functions are called once. 

\texttt{REQUESTINIT} and \texttt{REQUESTSHUTDOWN} are called every time the interpreter receives a request and finishes processing it, respectively. The former retrieves the HTTP request ID--assigned by the fuzzer--from the custom header of the HTTP request and sets the \texttt{coverage\_id} to it. When \texttt{coverage\_id} is not set, it indicates non-fuzzing request and all handlers simply return to avoid the instrumentation's overhead. This enables deployment on production servers where both developers (submit non-fuzzing requests) and security testers (submit fuzzing requests) work together. The de facto standard instrumentation, PCOV~\cite{krakjoepcov} and UOPZ~\cite{krakjoeuopz}, lacks this capability. \texttt{REQUESTINIT} also clears \texttt{path\_table} and initializes \texttt{f\_call\_sequence}, and \texttt{tainted\_branch\_\allowbreak log}. The latter writes the logs to the file system and frees them. 

\texttt{Error\&ExceptionLogger} has two separate handlers for error and exception, respectively, despite being combined at \circnum{6} for illustration. The handler for \texttt{ErrorObserver} is not shown for brevity. A \texttt{FunctionCall\allowbreak Handler} is defined to monitor the execution of vulnerable functions belonging to certain vulnerability class and to monitor sanitization functions. This strategy simplifies the architecture by providing a uniform monitoring technique to intercept all vulnerable function calls within a vulnerability class, without sacrificing modularity and extendability. The backward constant probe are invoked at \circnum{5} and \circnum{8} with \texttt{sink} (a function parameter that may be tainted by user input) and the \texttt{Frame} as parameters. All variables in the procedures are some of the attributes of the respective log (coverage, error and exception, function trace, and branch parameter log); other attributes are omitted for brevity.    

\begin{algorithm}[h!t] 
\small
\caption{Instrumentation core pipeline.}
\label{alg:instrcorepipe}
\begin{algorithmic}[1] 

\State \textbf{Module Global Variables:}
\State $coverage\_id$
\State $path\_table,\;f\_call\_sequence,\;tainted\_branch\_log$

\Procedure{MODULEINIT}{$ModuleName$}
    \State Install BranchOpcodesHooking\Comment{Coverage collector}
    \State Install ZendThrowHooking\Comment{Exception logger}
    \State Install LangConstructHooking\Comment{Include or eval}
    \State Install Error\&ExceptionHooking\Comment{Error\&Exception logger}
    \State Register ErrorObserver\Comment{Error logger}
    \State Install FunctionCallHooking\Comment{Function tracer} \circnum{1}
\EndProcedure

\Procedure{GLOBALSINIT}{$ModuleName$}
    \State Initialize $path\_table$ memory
\EndProcedure


\Procedure{REQUESTINIT}{$ModuleName$}
    \State $coverage\_id \gets$ coverage id held in x-header
    \State Clear $path\_table$
    \State Initialize $f\_call\_sequence, tainted\_branch\_log$  
\EndProcedure

\Procedure{REQUESTSHUTDOWN}{$ModuleName$}
    \State Write $path\_table$\Comment{coverage log} 
    \State Write $f\_call\_sequence$\Comment{function/construct trace} 
    \State Write $tainted\_branch\_log$\Comment{input-tainted branches} 
    \State Free $f\_call\_sequence, tainted\_branch\_log$
\EndProcedure

\Procedure{BranchOpcodesHandler}{$Frame$} \circnum{2} 
    \State $location \gets FilePath(Frame) + LineNo(Frame)$
    \State Select case matching $CurrentInstr(Frame).Opcode$
    \State $path\_condition \gets$ Truth value of CurrentInstr
    \State Append $location$ and $path\_condition$ to $path\_table$
    \State Identify $input$ in branch, append to $tainted\_branch\_log$ 
\EndProcedure

\Procedure{ZendThrowHandler}{$Frame$} \circnum{3} 
    \State $location \gets FilePath(Frame) + LineNo(Frame)$
    \State $opline \gets$ CurrentInstr(Frame) 
    \State Extract exception object from $opline$
    \State Log exception
\EndProcedure

\Procedure{LangConstructHandler}{$Frame$} \circnum{4} 
	\State $location \gets FilePath(Frame) + LineNo(Frame)$
    \State $opcode \gets CurrentInstr(Frame).Opcode$
    \State Select case matching $opcode$
    \State $ext\_opcode \gets$ Extended opcode from CurrentInstr
   	\State $params \gets$ Parameters, can be a \textbf{sink}
	\State $is\_const \gets$ IsStringLiteral($sink, Frame$) \circnum{5} 
	\State Call original handler
    \State Check success or error
    \State Append the trace to $f\_call\_sequence$ 
\EndProcedure

\Procedure{Error\&ExceptionLogger}{$Info$} \circnum{6} 
	\State $location \gets$ FilePath and Lineno from the Frame
	\State $f\_name \gets$ FunctionName from the Frame
	\State $message \gets$ $Info$
    \State Write error \& exception log to filesystem 
\EndProcedure

\Procedure{FunctionCallHandler}{$Frame$} \circnum{7} 
	\State $location \gets FilePath(Frame) + LineNo(Frame)$
	\State $f\_name \gets FunctionName(Frame)$
	\State $params \gets$ Parameters, some of which are \textbf{sinks}
	\State $is\_const \gets IsStringLiteral(sink, Frame)$ \circnum{8} 
	\State $return\_val \gets$ Call original handler
    \State Append the trace to $f\_call\_sequence$ 
\EndProcedure
\end{algorithmic}
\end{algorithm}

\subsubsection{Backward Constant Probe}

Algorithm~\ref{alg:backconstprober} outlines the internal process of the  backward constant probe (see Section~\ref{sec:originate}). It starts while the hook for the vulnerable function or language construct is active, so the sink is known in advance. It first retrieves the opcode array of the previous stack frame of the caller function--as shown in lines 2. The opcode array is a compile-time data structure containing a sequence of opcodes generated from a userland module. Following the definition of constant origination in Section~\ref{sec:originate}, it then iterates over the literals array to find a literal string that matches the parameter \texttt{Param} at \circnum{1}, and returns \texttt{1} if found. If not found, the probe continues with traversing down the opcode array to find an instruction whose opcode pushes \texttt{Param} to the stack: \texttt{SendVar} opcode at \circnum{2}. This instruction immediately precedes the instruction being hooked--an instruction that executes the internal function, making its identification very efficient. If the first operand of \texttt{SendVar} is a constant, the function returns 1. If not, \texttt{FromConst} function is called. The statement at \circnum{3} serves as a fallback mechanism, anticipating that different opcodes may be used to push parameters onto the stack in the future.

The \texttt{FromConst} function probes if an operand originates from a constant, by recursively walking down the opcode array. For efficiency, the statement at \circnum{4} limits the walk to the caller frame only. However, by making a few adjustments, including the statements at lines 2--3, the walk can extend to deeper frames. The probe leverages the semantics of a few string-manipulating opcodes, categorized as \texttt{ConstantOpcode} (fetch or declare), \texttt{Assignment}, and \texttt{Concate\allowbreak nation}, annotated as \circnum{5}, \circnum{6}, and \circnum{7}, respectively.


\begin{algorithm}[h!t] 
\small
\caption{Backward constant prober}
\label{alg:backconstprober}
\begin{algorithmic}[1] 

\Function{IsStringLiteral}{$Param$, $Frame$}
    \State $instrArray \gets OpcodeArray(Frame.PrevFrame.Function)$
    \For{\textbf{each} $literal$ \textbf{in} $instrArray.Literals$} \circnum{1}
    	\If{$Param = literal$} 
    		\State \Return $1$ \Comment{Param $\approx$ caller's compile-time literal}
		\EndIf
    \EndFor
    \State $i \gets CurrentInstr(Frame.PrevFrame)$ \Comment{Call instruction}
    \For{$i$ \textbf{downto} $firstInstruction$ \textbf{in} $instrArray$}
   		\If{$(i.Opcode$ \textbf{is} $SendVar) \land (i.Op1$ \textbf{is} $Param)$} \circnum{2}
   			\If{$i.Op1$ \textbf{is} $Const$} \Comment{Op1 is the $1^{\text{st}}$ operand}
    				\State \Return$1$
    		\EndIf
    		\State \Return{FromConst($i.Op1$,$i$,$instrArray$)}
		\EndIf
   	\EndFor
	\State \Return{$0$} \circnum{3}\Comment{fallback, if $SendVar$ changes}
\EndFunction

\Statex
\Function{FromConst}{$Operand$, $I$, $InstrArray$}
	\If{$I \leq InstrArray.FirstInstruction$}
		\State \Return $0$ \circnum{4}	
	\EndIf
	\State $I \gets I-1$
	\If{$I.Result = Operand$}
		\If{$I.Opcode$ \textbf{is} $ConstantOpcode$} \circnum{5}
			\State \Return $1$
		\ElsIf{$I.Opcode$ \textbf{is} $Assignment$} \circnum{6} \Comment{// Op1 = Op2}
			\If{$I.Op2.Type$ \textbf{is} $Const$}
				\State \Return $1$
			\EndIf
            \State \Return{FromConst($I.Op2$,$I$,$InstrArray$)}
		\ElsIf{$I.Opcode$ \textbf{is} $Concatenation$} \circnum{7} \Comment{// Op1 . Op2}
			\State $op1Const \gets I.Op1.Type$ \textbf{is} $Const$
			\State $op2Const \gets I.Op2.Type$ \textbf{is} $Const$
			\If{$op1Const \land op2Const$}
				\State \Return $1$
			\EndIf
			\If{$!op1Const \land (I.Op1.Type \neq Unused) \land op2Const$}
				\State \Return{FromConst($I.Op1$,$I$,$InstrArray$)}
			\EndIf
			\If{$op1Const \land !op2Const \land (I.Op2.Type \neq Unused)$}
				\State \Return{FromConst($I.Op2$,$I$,$InstrArray$)}
			\EndIf
			\State \Return{$0$}
		\ElsIf{$!I.Result \land (I.Op1.Type$ \textbf{is} $Operand$)}
			\If{$I.Op2.Type$ \textbf{is} $Const$}
				\State \Return{$1$}
			\ElsIf{$I.Op2.Type \neq Unused$}
				\State \Return{FromConst($I.Op2$,$I$,$InstrArray$)}
			\EndIf
			\State \Return{$0$}
		\EndIf
		\State \Return{FromConst($Operand, I, InstrArray$)}  
	\EndIf
\EndFunction
\end{algorithmic}
\end{algorithm}

\subsubsection{Trace Analyzer}

Algorithm~\ref{alg:traceanalyzer} outlines the internal process of the trace analyzer. The fuzzer lazily calls \texttt{SetCallTrace} function in line 1, only when a new path is found--indicating a new input-- to collect vulnerable and sanitization function semantics. The \texttt{SetCallTrace} function receives the input object--a dictionary--and a list of hashes of vulnerable function traces (\texttt{VulnFuncHashes}). We extend Phuzz's input class (\texttt{class Candidate}) to store several fields—including trace status, function traces, a list of sanitization function traces, a list of vulnerable function traces, the path hash, and a list of completed fuzzing targets—where the first four are used in Algorithm~\ref{alg:traceanalyzer} (lines 2, 3, 11, and 25).

The \texttt{SetCallTrace} first identifies \texttt{die} or \texttt{exit} functions--indicating premature script termination--and collects the traces of sanitization and vulnerable functions by iterating over the function traces, at \circnum{1}, \circnum{2}, \circnum{3}, respectively. It also filters out vulnerable functions whose sink originates from constant values, denoted by \circnum{4}. It then checks if a trace of vulnerable function is new~\circnum{5}. If so, the trace is added the list, making it a fuzzing target and calls \texttt{ParamsInSink} \circnum{6} to identify inputs that taint the sinks (the function parameters).

The \texttt{ParamsInSink} function identifies inputs that taint the sink, allowing function-level fuzzing to focus mutations exclusively on the tainted parameters. This function loops in the function parameters (\texttt{params}) and matches each parameter's value to a sink via the \texttt{PayloadInSink} substring matching heuristic \circnum{7}, avoiding pre-fuzzing taint analysis overhead incurred by~\cite{zhao2022,wang2025}. If matched, the parameter key is added to \texttt{pInSink} and its data type and quoted flag are set to \texttt{None}. Further, when the trace belongs to a function vulnerable to SQLI, the data type and quoted flag are determined by parsing the query at \circnum{8} and retrieving the database metadata at \circnum{9}. Finally, if the result of parsing matches the metadata annotated by \circnum{10}, the data type and quoted flag are assigned.

\begin{algorithm}[h!t] 
\small
\caption{Trace analyzer}
\label{alg:traceanalyzer}
\begin{algorithmic}[1] 

\Procedure{SetCallTrace}{$in$, $VulnFuncHashes$}
	\State $status \gets 0$ 
    \State $traces \gets in.FuncTraces$ \Comment{traces is a list of dictionaries}
    \For{ $f$ \textbf{in} $traces$}\Comment{f is a dictionary} 
    		\State $isVuln \gets False$
    		\State $name \gets f[name]$
    		\If{$name \in DieExitFunctions$} \circnum{1}
    			\State $status \oreq ExistDieExit$ 
    		\ElsIf{$name \in SanitizeFunctions$} \circnum{2}
    			\State $status \oreq ExistSanitation$
    			\State $in.sanitations.append(f)$	
    		\ElsIf{$name \in VulnFunctions$} \circnum{3}
    			\State $status \oreq ExistVulnFunction$
    			\If{$f[sink] \neq Constant$} \circnum{4}
    				\State $isVuln \gets True$
    			\EndIf	
		\EndIf
		\If{$isVuln$}
			\State $hash \gets$ GetHash($f$)
			\If{$hash \notin VulnFuncHashes$} \circnum{5}
				\If{$f \notin in.VulnFunctions$}
					\State $VulnFuncHashes.add(hash)$
					\State $f[Status] \gets FUZZED$
					\State $f[funcIteration] \gets 0$
					\State $f[paramIteration] \gets 0$
					\State $f[pInSink] \gets$ ParamsInSink($in.p$, $sink$, $f$) \circnum{6}
					\State $in.VulnFunctions.append(f)$
				\EndIf
			\EndIf
		\EndIf
    \EndFor
    \State $in[Status] \gets status$
\EndProcedure
\Statex
\Function{ParamsInSink}{$params$, $sink$, $functionTrace$}
	\State $f \gets  functionTrace$
	\State $pInSink \gets \{\}$\Comment{a dictionary}
	\For{ $key, value$ \textbf{in} $params$}\Comment{params is a dictionary}
		\If{$PayloadInSink(value,sink)$} \circnum{7}
			\State $pInSink[key] \gets$ $\{type:None,quoted:None\}$ 
			\If{$f \in SQLIVulnFunctions$}
				\State $tableSet \gets$ parse($f[query]$) \circnum{8}
				\State $desc \gets$ Metadata($tableSet$,$DBconn$) \circnum{9}
				\If{$tableSet$ \textbf{matches} $desc$} \circnum{10}
				\State $pInSink[key][type] \gets$ $desc[type]$
				\State $pInSink[key][quoted] \gets$ $desc[quoted]$
				\EndIf		
			\EndIf	
		\EndIf
	\EndFor
	\Return{$pInSink$}
\EndFunction
\end{algorithmic}
\end{algorithm}

\section{Implementation}

We implement the back-end instrumentation in C, comprising 5.4k LOC, and deploy it as the PHP extension \toolInstr{} with support to PHP 8.3 and 8.4. \toolFuzz{}, the redesigned version of the Phuzz \cite{neef2024} fuzzer, are implemented in 4.9k LOC Python. The Algorithm~\ref{alg:instrcorepipe} initialization and shutdown functions are directly implemented in Zend's lifecycle functions: \texttt{MINIT}, \texttt{GINIT}, and \texttt{RINIT} for initialization and \texttt{MSHUTDOWN}, \texttt{GSHUTDOWN} and \texttt{RSHUTDOWN} for shutdown. To represent the program's branch coverage we provide hooks for 16 branch opcodes (Appendix~\ref{sec:appendix} Table~\ref{tab:branchops}).


We also hook \texttt{ZEND\_THROW} for reporting exceptions and three opcodes for language constructs (Appendix~\ref{sec:appendix} Tables~\ref{tab:zendthrow} and~\ref{tab:langconstops}). We implement 117~hooks for userland functions consisting of 74~potentially vulnerable functions and 36~sanitization functions (Appendix~\ref{sec:appendix} Table~\ref{tab:hookedfuncs}). The remaining 7~functions are MySQL bind and execute functions. The function and language construct tracer, input-tainted branch identifier, and error and exception logger uses the cJSON library~\cite{gamble2025} to store logs in the JSON format. For efficiency, ZIMPAF buffers all logs (except for errors and exceptions) per HTTP request, writing them in batch during RSHUTDOWN. We log errors and exceptions immediately upon interception to prevent trace loss if the Zend interpreter terminates abruptly due to a fatal error. This incurs minimal overhead due to their small size.

\toolFuzz{} prioritizes and scores inputs based on function trace semantics, namely the existence of vulnerable functions and language constructs, sanitization functions, and die or exit functions. \toolFuzz{} does not write mutations executing known paths to disk, using mutation hashes instead. To distinguish mutations, it appends the iteration count to the coverage ID, yielding a unique identifier for each mutation while saving disk space. This enables efficient state restoration if fuzzing terminates prematurely and resumes in the next run. Finally, the \toolFuzz{}'s trace analyzer uses the SQLGlot library~\cite{mao2023} to parse SQL queries into an abstract syntax tree (AST).

\section{Evaluation}

To evaluate \toolInstr{}+\toolFuzz{} we ask the following research questions~(RQs).
\begin{itemize}
    \item RQ1: Can \toolInstr{}+\toolFuzz{} find more vulnerabilities than the current state-of-the-art Phuzz?
    \item RQ2: How does the runtime of \toolInstr{}+\toolFuzz{} compare to the state-of-the-art Phuzz?
\end{itemize}    

We compare the \toolInstr{} and \toolFuzz{} framework with the state-of-the-art Phuzz~\cite{neef2024} for two reasons. First, among the state-of-the-art fuzzers~\cite{vanRooij2021,zhao2022,trickel2023,guler2024,wang2025}, Phuzz provides the strongest vulnerability-detection baseline on overlapping web applications and vulnerability classes, as derived from the published evaluations of Phuzz, Atropos~\cite{guler2024}, Predator~\cite{wang2025}, and Witcher~\cite{trickel2023}. 
Second, \toolInstr{} runs on PHP $\ge$ 8.3 and is directly comparable to Phuzz's instrumentation (PCOV~\cite{krakjoepcov} and UOPZ~\cite{krakjoeuopz} extension) that can be installed on the same PHP version. In contrast, other state-of-the-art fuzzers~\cite{trickel2023,guler2024,wang2025} rely on a patched runtime instrumentation restricted to PHP 7.x, which has reached end-of-life. Major architectural overhauls between PHP 7 and PHP 8 (e.g., JIT compilation, redesigned memory allocation, and ZVAL tracking) substantially alter VM performance. Consequently, a direct quantitative comparison would require re-engineering those frameworks for PHP~$\geq8.3$, rendering cross-version benchmarks invalid. 

The evaluation benchmark consists of a PoC web application containing code for Listings~\ref{lst:listing-1} and~\ref{lst:listing-2} and five web applications from Phuzz's benchmark. Phuzz's benchmark comprises four intentionally vulnerable web applications~\cite{bwapp2022,dvwa2020,doupe2010,xvwa2015} and one commercial web application, Wordpress (``WP''), comprising 17 plugins. Most of these plugins are big or medium. For instance, Arprice-responsive-pricing-table-3.6 executes $\ge$35,000 functions and language construct, and $\ge$81,000 branch instructions, producing $\ge$29 MB logs. Furthermore, we create HTTP request templates for the PoC web application, whereas other benchmarks rely on Phuzz-provided templates to generate fuzzing inputs (see Figure~\ref{fig:architecture}).

Both web server and the fuzzer run on a 4~core 1.8~GHz Intel Core i7-8550U CPU machine with 16 GB of RAM and NVMe SSD 2 TB. The benchmarks are hosted on Ubuntu with Apache~2.4.58 \texttt{mod\_php} and PHP 8.3.19 with \texttt{OPcache} extension~\cite{phpnet} enabled for high speed request processing by avoiding script recompilation.  

\textbf{RQ1:} Table~\ref{tab:performance-comparison} compares the performance of \toolInstr{}+\toolFuzz{} with Phuzz (summarizing Appendix~\ref{sec:appendix} Tables~\ref{tab:eval_poc_wordpress},~\ref{tab:eval_dvwa},~\ref{tab:eval_bwapp},and~\ref{tab:eval_wacko_xvwa}). Six impossible test cases from DVWA are omitted since they are not vulnerable (see Table~\ref{tab:eval_dvwa}). Phuzz misses 16 vulnerabilities, including seven vulnerabilities arising from language constructs that its instrumentation (\texttt{UOPZ})~\cite{krakjoeuopz} cannot intercept, requiring manual inspection. Rows with bold numbers under Phuzz's \texttt{\#detections} column represent benchmarks where Phuzz misses vulnerabilities. Conversely, our framework detects all vulnerabilities; however, it reports one as a bug and produces one false positive. Overall, the framework is faster (21\%) despite detecting more vulnerabilities and 2$\times$ faster when considering fully overlapped vulnerabilities (excluding rows with bold numbers). On per test case basis, the framework is 77\% faster, including the vulnerabilities undetected by Phuzz. The framework also uncovers more than 2$\times$ number of paths than Phuzz as shown by \texttt{\#new paths} columns of Table~\ref{tab:performance-comparison}. The \#vf\textsubscript{skip} column demonstrates that the backward constant probe identifies 10,458 untainted functions, enabling RedPhuzz to skip them and avoid massive computational overhead and execution time.

\begin{table*}[!t]
\centering
\caption{Comparison of ZIMPAF \& RedPhuzz framework and Phuzz; 
vulns~=~vulnerabilities detected; 
+paths~=~paths newly discovered; 
t~=~average detection time; 
vf\textsubscript{skip}~=~untainted vulnerable functions identified via backward constant probe and skipped during function-level fuzzing;
faster~=~vulns where framework detects faster;
bold~=~more vulnerabilities discovered or runtime reduced over 50\%.
}
\label{tab:performance-comparison}
\small
\begin{tabular}{lrrrrrrrrr}
\toprule
&
\multicolumn{4}{c}{Phuzz} & \multicolumn{5}{c}{ZIMPAF \& RedPhuzz}\\
\cmidrule(lr){2-5} \cmidrule(lr){6-10}
Benchmark (vulns) &
vulns &
+paths &
t (s) &
faster &
vulns &
+paths &
t (s) &
vf\textsubscript{skip} &
faster \\
\midrule

\multicolumn{8}{l}{PoC (3)} \\
\hspace{0.8em}1$\times$ Path traversal
    & 0 & 0 & -- & 0 & \textsuperscript{\dag} \textbf{1} & 1 & 1.0 & 3 & 1 \\
\hspace{0.8em}2$\times$ SQL injection
    & 0 & 0 & -- & 0 & \textbf{2} & 2 & 0.7 & 5 & 2 \\

\multicolumn{8}{l}{WP Plugins (17)} \\
\hspace{0.8em}2$\times$ Path traversal
    & 1 & 1 & 17.0 & 1 & \textbf{2} & 3 & 44.4 & 810 & 1\\
\hspace{0.8em}7$\times$ SQL injection
    & 7 & 13 & 97.1 & 3 & 7 & 38 & 70.2 & 5,037 & 4 \\
\hspace{0.8em}6$\times$ XSS injection
    & 6 & 15 & 107.0 & 3 & 6 & 19 & 104.7 & 3,259 & 3 \\
\hspace{0.8em}2$\times$ Deserialization 
    & 0 & 3 & -- & 0 & \textsuperscript{\ddag} \textbf{2} & 3 & 88.3 & 824 & 2 \\

\multicolumn{8}{l}{DVWA (17)} \\
\hspace{0.8em}3$\times$ Path traversal
    & 0 & 1 & -- & 0 & \textbf{3} & 2 & 1.0 & 15 & 3 \\
\hspace{0.8em}3$\times$ Remote exec
    & 3 & 0 & 4.5 & 0 & 3 & 0 & 3.4 & 18 & 3 \\
\hspace{0.8em}5$\times$ SQL injection
    & 5 & 12 & 38.1 & 0 & 5 & 6 & \textbf{1.2} & 32 & 4 \\
\hspace{0.8em}6$\times$ XSS injection
    & 6 & 0 & 11.6 & 2 & 6 & 4 & \textbf{3.6} & 42 & 4 \\

\multicolumn{8}{l}{bWAPP (17)} \\
\hspace{0.8em}2$\times$ Path traversal
    & 0 & 0 & -- & 0 & \textbf{2} & 1 & 0.8 & 19 & 2 \\
\hspace{0.8em}3$\times$ Remote exec
    & 2 & 1 & 3.4  & 0 &\textbf{3} & 2 & \textbf{1.2} & 27 & 3 \\
\hspace{0.8em}9$\times$ SQL injection
    & 9 & 8 & 7.0 & 2 & 9 & 10 & \textbf{0.6} & 117 & 7 \\
\hspace{0.8em}11$\times$ XSS injection
    & 11 & 0 & 79.1 & 2 & 11 & 6 & \textbf{0.7} & 115 & 9 \\

\multicolumn{8}{l}{WackoPicko (9)} \\
\hspace{0.8em}2$\times$ Path traversal
    & 0 & 0 & -- & 0 & \textbf{2} & 8 & 8.6 & 11 & 2 \\
\hspace{0.8em}1$\times$ Remote exec
    & 1 & 1 & 14.2 & 0 & 1 & 1 & \textbf{0.1} & 6 & 1 \\
\hspace{0.8em}2$\times$ SQL injection
    & 1 & 2 & 24.2 & 0 & \textbf{2} & 4 & \textbf{0.8} & 24 & 2\\
\hspace{0.8em}4$\times$ XSS injection
    & 4 & 2 & 13.8 & 1 & 4 & 3 & \textbf{4.8} & 35 & 3\\

\multicolumn{8}{l}{XVWA (17)} \\
\hspace{0.8em}1$\times$ Path traversal
    & 0 & 0 & -- & 0 & \textbf{1} & 0 & 0.3 & 6 & 1 \\
\hspace{0.8em}1$\times$ Remote exec
    & 1 & 0 & 2.7 & 0 & 1 & 1 & \textbf{0.1} & 6 & 1 \\
\hspace{0.8em}4$\times$ SQL injection
    & 4 & 1 & 1.6 & 2 & 4 & 9 & \textbf{0.5} & 28 & 2 \\
\hspace{0.8em}1$\times$ Deserialization
    & 1 & 0 & 0.8 & 0 & 1 & 0 & 0.6 & 6 & 1 \\
\hspace{0.8em}2$\times$ XSS injection
    & 2 & 0 & 4.1 & 1 & 2 & 3 & \textbf{0.9} & 13 & 1 \\

\midrule
Total (80)
    & 64 & 60 & 426.1 & 17 & \textbf{80} & 126 & 338.6 & 10,458 & 62  \\

fully overlapped vulns
    & & & 381.5 &  & & & 190.4 & &  \\
    
\bottomrule
\end{tabular}%
\vspace{2pt}
 
\makebox[\textwidth][l]{%
    \footnotesize \textsuperscript{\dag} reported as a bug. 
    \footnotesize \textsuperscript{\ddag} RedPhuzz reports additional Path Travs. vulnerability due to file permission error, a false positive.
}
\end{table*}

As shown in Table~\ref{tab:performance-comparison}, Phuzz fails to identify any vulnerabilities and bug in the PoC web application because its lackof execution-context awareness blinds it to the applied sanitization, input-tainted branch, tainted sink in vulnerable functions, rendering its mutation strategy ineffective. Conversely, equipped with this awareness, \toolFuzz{} effectively mutates parameters to reach vulnerable functions and trigger vulnerabilities or bugs. Other benchmarks further confirm Phuzz's limitation as summarized in Table~\ref{tab:wordpress_vuln_func},~\ref{tab:bwapp_vuln_func}, and~\ref{tab:wackopicko_vuln_func}.

\phantomsection
\label{sec:exploitability}
Furthermore, Phuzz incorrectly classifies Test No. 46 as vulnerable (false positive), as reported in Table~\ref{tab:eval_bwapp}. As shown in Table~\ref{tab:bwapp_vuln_func}, Phuzz logs an error from \texttt{is\_file} when given an invalid path, although this function is not vulnerable but instead guards \texttt{fopen}, the actual vulnerable function. In contrast, sanitization-aware mutation enables \toolFuzz{} to generate a valid, malicious path that reaches \texttt{fopen}. The function-trace-based checker then identifies the vulnerability when the malicious, well-formed path reaches \texttt{fopen} and its return value indicates successful execution. This case demonstrates \toolFuzz{}'s ability to identify silent vulnerabilities based on exploitability.









\begin{table*}[h!t]
\caption{Throughput of \toolInstr{} and PCOV+UOPZ based on Table~\ref{tab:write-details-time} averaged over 10 sample; 
w~=~write volume;
tp~=~throughput; \\
scale~=~ZIMPAF's tp / (PCOV + UOPZ) tp;
arprice~=~ Arprice-responsive-pricing-table 3.6 WordPress plugin;
bold~=~at least 2$\times$ time or throughput improvement.
}
\label{tab:throughput}
\centering
\small
\begin{tabular}{lrrrrrrr}
\toprule
& \multicolumn{3}{c}{PCOV + UOPZ}  & \multicolumn{3}{c}{\toolInstr{}} & \\
\cmidrule(lr){2-4} \cmidrule(lr){5-7} \cmidrule(lr){8-8}
& w (kB) & t (ms) & tp (B/ms) & w (kB) & t (ms) & tp (B/ms) & scale \\
\midrule
\texttt{arprice} & 5,802.9 & 6,553.5 & 885 & 29,476.3 & 4,166.9 & \textbf{7,073} & 8.0 \\
\texttt{bwapp\_sqli\_ajax} & 2.9 & 12.9 & 224 & 2.2 & 11.0 & 195 & 0.9 \\
\texttt{dvwa\_sqli\_low} & 17.9 & 678.5 & 26 & 4.6 & \textbf{4.2} & \textbf{1,088} & 41.2 \\
\texttt{poc\_seq\_sanit} & 0.4 & 6.0 & 68 & 1.4 & 6.2 & \textbf{225} & 3.3 \\
\texttt{wacko\_register\_1} & 2.3 & 10.7 & 214 & 4.2 & 8.4 & \textbf{506} & 2.4 \\
\texttt{xvwa\_sqli\_search} & 1.3 & 14.7 & 91 & 2.3 & 12.1 & 191 & 2.1 \\
\bottomrule
\end{tabular}
\end{table*}

\textbf{RQ2:} The superior capability above is enabled by \toolInstr{}'s efficient, fine-grained interpreter instrumentation, which provides high-fidelity runtime semantics feedback to guide \toolFuzz{}'s fuzzing. Unlike PCOV and UOPZ, which primarily provide coverage, error, and exception feedback, \toolInstr{} produces richer runtime traces with substantially lower overhead, as shown in Table~\ref{tab:throughput}. The table reports instrumentation time and data volume for the first request of each sampled benchmark, ensuring comparability because the execution paths are identical across tools. \toolInstr{} achieves 2.1--41.22$\times$ higher throughput than PCOV and UOPZ on five benchmarks, and 0.87$\times$ on one. Table~\ref{tab:tracebranch} further illustrates \toolInstr{} workload by detailing the instrumentation for the first entry (arprice) in Table~\ref{tab:throughput}. It demonstrates \toolFuzz{}'s strong semantic filtering capability to identify 9 vulnerable functions for fuzzing target and 15 applied sanitization.  

\begin{table}[t]
\caption{Arprice-responsive-pricing-table-3.6 trace and tainted-branch log details for the first request of the original (unmutated) input}
\label{tab:tracebranch}
\centering
\small
\begin{tabular}{p{0.72\columnwidth}r}
\toprule
Metric & Value \\
\midrule
size of function call and construct traces log & 22,804.5 kB \\
intercepted function calls \& construct executions & 35,623 \\
\hspace{7pt}- vulnerable functions \& constructs & 2,339 \\
\hspace{7pt}- untainted vuln. functions \& constructs \circnum{1} & 914 \\
\hspace{7pt}- traversal constructs with codebase path \circnum{2} & 996 \\
\hspace{7pt}- duplicate vuln. functions/ constructs removed & 420 \\
\hspace{7pt}- input's vuln. functions \& constructs (targets) & 9 \\
\midrule
sanitization functions (duplicate removed) & 27,251 \\
sanitization applied to input parameters & 15 \\
\midrule
intercepted branches & 81,730 \\
size of branch parameters log & 2.1 kB \\
entries in branch parameters log & 7 \\
distinct input-tainted branches & 4 \\
\bottomrule
\end{tabular}
\end{table}

\section{Related Work}
\label{sec:related-works}

Grey-box fuzzing addresses the limitations of black-box fuzzing~\cite{burp2026, checkmarx2026, acunetix2026, doupe2012, pellegrino2014}, which lacks direct visibility into the internals of applications and therefore limits path exploration. It also mitigates inefficiency of white-box fuzzing~\cite{artzi2010,alhuzali2018}, with full visibility, suffers from path explosion and expensive constraint solving. To realize a grey-box fuzzer, American Fuzzy Lop (AFL)~\cite{zalewsky} instruments native applications to collect code coverage--runtime execution semantics representing the application internals--as feedback to guide fuzzing. Mutations producing unique coverage are added to the corpus and ranked, allowing AFL to prioritize impactful inputs and explore diverse paths for potential bugs or vulnerabilities.

The AFL approach has also been adopted to improve the fuzzing of server-side web applications.

The most related state-of-the-art grey-box fuzzers are discussed in Section~\ref{sec:intro}, forming the basis of our arguments for proposing \toolInstr{} and \toolFuzz{} to explain their weaknesses, and in Section~\ref{sec:design} to explain how our design addresses them. Here, we position our research in the broader grey-box fuzzing landscape. Grey-box fuzzing resolves the lack of feedback and limited mutation capability of black-box fuzzers which impair their effectiveness, such as Burp Suite and ZAP~\cite{burp2026,checkmarx2026}. Similarly, it mitigates the typically high computational overhead of static analysis~\cite{dahse2014,luo2022} and white-box fuzzing~\cite{huang2019}, which often produces many false positives. 

WDFUZZ employs static analysis to provide fuzzing inputs, which are then fuzzed to uncover vulnerability and explore new path~\cite{lin2025}. \toolFuzz{} differs in that inputs are known in advance, provided as JSON files, most of which inherited from Phuzz, indicating that the entire fuzzing workflow is not fully seamless. This is an opportunity to extend \toolFuzz{} by integrating automatic input submission to \toolFuzz{} by only giving it an endpoint to fuzz. While WDFUZZ identifies sources and sinks statically and applied taint analysis to track tainted sinks, \toolInstr{} is aware of the sink when initiating backward constant probe (see Algorithm~\ref{alg:backconstprober} at \circnum{8}), and \toolFuzz{} directly identifies tainted sink using trace information (see Algorithm~\ref{alg:traceanalyzer} at \circnum{6}).

CorbFuzz, a browser fuzzer, also employs branch interception to provide coverage bitmap~\cite{shou2021}. However, it misses \texttt{ZEND\_JMP*} opcode that may cause incomplete coverage. Additionally, it installs the hooks in \texttt{RINIT}, resulting in per-request installation, which is inefficient. \toolInstr{} hooks the opcode and installs hooks once for all requests in \texttt{MINIT}. CorbFuzz also assumes that manually setting up a database is impractical, so it synthesizes a database and performs type inference automatically. \toolFuzz{} leverages an existing database to retrieve metadata for identifying data types for mutation. 

Spider-Scents detects XSS by injecting payloads directly into the database, bypassing the web application, and then crawling the webpage for XSS reflections~\cite{olsson2024}. Then, the human tester manually analyses the crawler output to determine whether the payloads can be injected via web application. This approach is inefficient. In contrast, \toolFuzz{} injects payloads upon encountering \texttt{INSERT} or \texttt{UPDATE} queries, enabling in‑situ fuzzing of stored XSS, and automated detection via the webFuzz~\cite{vanRooij2021} detector, which is a highly targeted and efficient strategy.

\section{Future Work}

We are currently optimizing \toolInstr{}, e.g., to manage the input growth given in Table~\ref{tab:performance-comparison}, allowing to enhance its capability, such as combining symbolic execution. As discussed in Section~\ref{sec:related-works}, we also plan to build automatic input generator for our framework capable of adapting with different web architecture. Future work could extend support for more sanitization-aware mutation as it is tailored to particular sanitization functions, such as adding preg\_replace which is absent in the existing mutation. It can also include building an SQL query parser to mitigate the failure of SQLGlot~\cite{mao2023} to parse complex queries containing malformed parameters, causing \toolFuzz{} to proceed without knowledge of the data type of query parameters and their quoted flag. Likewise, Python's \texttt{xeger} may not work when the regex pattern is complex, potentially leading to deep or infinite recursion that causes Python to terminate, even though the call is enclosed in \texttt{try}--\texttt{except} block (see Figure~\ref{fig:xeger_failure}). Finally, future work could target the detection of more vulnerability classes.

\section{Conclusions}

In this paper, we discuss several weaknesses in the state of the art fuzzers~\cite{vanRooij2021,zhao2022,trickel2023,guler2024, neef2024,wang2025} and how they limit their performance. We addresses the weaknesses by introducing a novel fuzzing framework consisting of \toolInstr{}, a runtime interpreter instrumentation and \toolFuzz{}, the front-end fuzzer. \toolInstr{} utilizes several novel techniques to achieve an efficient and effective instrumentation capable of producing information that is used by \toolFuzz{} to achieve multi-level and highly targeted fuzzing. The evaluation shows that our framework outperforms Phuzz.

\section{Acknowledgments}
The authors used ChatGPT and Gemini to check the paper's grammar and lightly polish the paper.

\bibliographystyle{plainurl}
\bibliography{ref}

\appendix

\section{Open Science}
\label{sec:open-science}

To support open science, we publish all framework artifacts comprising the source code for the instrumentation, the front-end fuzzer, and the benchmark web applications used for the evaluation, including our own proof of concept. The framework is downloadable at this link: \url{https://github.com/tennovs/zimpaf_redphuzz}.

The GitHub site also explains how to build and execute the evaluation environment via Docker, which resembles the one on our host machine for easy reproducibility. For efficient execution we suggest setting up the environment directly on host due to Docker's limitation. In this way, we expect to foster further research aiming to extend or improve our framework and to promote its wide adoption.

\section{Ethical Considerations}

The sole motivation of this research is to advance grey-box web application fuzzing, which has lagged behind similar research for native application, and to offer more tools for stakeholders to develop more secure web applications. The development of framework posed no risks since it was performed on local machine without any interactions with external systems. The benchmark consists of a proof-of-concept web application and several web applications with known or published vulnerabilities, the latter of which use inputs provided in the previous research. 

Our framework is released under Apache 2.0 license, enabling the stakeholders to benefit according to their purposes. The security community benefits from our research through the invention of novel techniques that can be further extended or improved by the community. Practitioners can use or integrate our framework into their workflows to find vulnerabilities and bugs, which results in more secure web applications. However, malicious users can also misuse our framework to uncover vulnerabilities and exploit them for harmful intents. We mitigate this risk by deciding to publish at this stage of our research, before full automation of input generation and endpoint identification is included in the framework. While vulnerabilities are discovered, neither we discuss nor give any hints on how to further escalate the detection into system compromised.

\section{Appendix}
\label{sec:appendix}

\begin{table}[h!t]
\small
\centering
\caption{Branch opcodes (each following a ZEND\_ prefix) for execution paths}
\label{tab:branchops}
\begin{tabularx}{\columnwidth}{X}
\toprule
JMP, JMPZ, JMPNZ, JMPZ\_EX, JMPNZ\_EX, JMP\_NULL, JMP\_SET, COALESCE, IS\_EQUAL, IS\_NOT\_EQUAL, IS\_IDENTICAL, IS\_NOT\_IDENTICAL, IS\_SMALLER, IS\_SMALLER\_OR\_EQUAL, CASE, CASE\_STRICT \\
\bottomrule
\end{tabularx}
\end{table}

\vspace{1.5em} 
\noindent 
\begin{minipage}{\columnwidth}
\small
\captionof{table}{Exception Opcode}
\label{tab:zendthrow}
\begin{tabular}{p{0.25\columnwidth}|p{0.68\columnwidth}}
\toprule
Opcode & Purpose \\
\midrule
\texttt{ZEND\_THROW} &
The last resort for catching an exception when the Zend interpreter bails out, bypassing normal execution flow, including \texttt{zend\_throw\_exception}. \\
\bottomrule
\end{tabular}
\end{minipage}

\vspace{1.5em} 
\noindent 
\begin{minipage}{\columnwidth}
\small
\captionof{table}{Opcodes (each following a ZEND\_ prefix) for language construct traces.
}
\label{tab:langconstops}
\begin{tabular}{p{.15\columnwidth}p{.2\columnwidth}p{.55\columnwidth}}
\toprule
Vuln. & Construct & Opcode [, Extended Value] \\
\midrule
\multirow{4}{*}{\parbox[t]{.15\columnwidth}{Path traversal}} 
    & include       & INCLUDE\_OR\_EVAL, INCLUDE \\
    & include\_once & INCLUDE\_OR\_EVAL, INCLUDE\_ONCE \\
    & require       & INCLUDE\_OR\_EVAL, REQUIRE \\
    & require\_once & INCLUDE\_OR\_EVAL, REQUIRE\_ONCE \\
Code Exec & eval & INCLUDE\_OR\_EVAL, EVAL \\
\midrule
- & exit, die & EXIT \\
\bottomrule
\end{tabular}
\end{minipage}

\vspace{2pt} 
\begin{minipage}[t]{7.5cm}
\small 
Extended value is a member of data structure (\texttt{\_zend\_op} struct) of a polymorphic language construct opcode. It serves as an integer flag to identify the sub-type of the opcode being executed and to select the corresponding sub-operation (backpatching). \\
\end{minipage}

\begin{minipage}{\columnwidth}
\small
\captionof{table}{Functions being Monitored}
\label{tab:hookedfuncs}
\begin{tabular}{p{0.13\columnwidth}p{0.67\columnwidth}}
\toprule
Code Execution
& assert, system, exec, passthru, shell\_exec, popen, proc\_open \\
\midrule

Path traversal
& chgrp, chown, chmod, copy, delete, file, file\_get\_contents, fopen, glob, lchgrp, lchown, link, mkdir, move\_uploaded\_file, parse\_ini\_file, parse\_ini\_string, readfile, rename, rmdir, stat, symlink, tempnam, touch, unlink, scandir, header, clearstatcache, disk\_free\_space, disk\_total\_space, file\_get\_contents, fileatime, filectime, filegroup, fileinode, filemtime, 
fileowner, filesize, fileperms, filetype, linkinfo, lstat, readlink \\
\midrule

MySQL Injection 
& mysqli::query, mysqli::execute\_query, mysqli::real\_query, mysqli::multi\_query, mysqli::prepare, PDO::query, PDO::prepare, PDO::exec, mysqli\_query, mysqli\_execute\_query, mysqli\_real\_query, mysqli\_multi\_query, mysqli\_prepare \\
\midrule

Deseriali \allowbreak zation
& unserialize, yaml\_parse, yaml\_parse\_file, igbinary\_unserialize \\
\midrule

XXE
& DOMDocument::load, DOMDocument::loadXML, XMLReader::XML, XMLReader::open, XMLReader::read, simplexml\_load\_string, simplexml\_load\_file, xml\_parse \\
\midrule

Sanitize
& mysqli\_real\_escape\_string, mysqli::real\_escape\_string, PDO::quote, htmlspecialchars, htmlentities, addslashes, stripslashes, strip\_tags, preg\_replace, preg\_match, realpath, basename, escapeshellarg, escapeshellcmd, str\_replace, strpos, stripos, filter\_var, filter\_input, filter\_var\_array, filter\_input\_array, libxml\_disable\_entity\_loader, is\_numeric, base64\_decode, json\_decode, fnmatch, is\_file, file\_exists, is\_dir, is\_excutable, is\_link, is\_readable, is\_writable, is\_uploaded\_file, dirname, pathinfo \\
\midrule

MySQL (bind|exec)
& mysqli\_stmt\_bind\_param, mysqli\_stmt\_execute, mysqli\_stmt::bind\_param, mysqli\_stmt::execute, PDOStatement::bindParam, PDOStatement::bindValue, PDOStatement::execute \\
\bottomrule
\end{tabular}
\end{minipage}

\begin{table*}[t]
\caption{Evaluation of \toolFuzz{} and Phuzz on PoC Web App (top) and Wordpress Plugins (bottom); p~=~new paths, t(s)= detection time, vf\_skip~=~ the number of untainted vulnerable functions identified by the backward constant probe and skipped during function-level fuzzing.
}
\label{tab:eval_poc_wordpress}
\centering
\small
\begin{tabular}{p{3.2cm}cp{2.3cm}|rrr|rr|p{4cm}}
\toprule
Benchmark &
Test &
Bug / &
\multicolumn{3}{c|}{\toolFuzz{}} & 
\multicolumn{2}{c|}{Phuzz} &
Info \\
&
No. &
Vulnerability &
t (s) & p & vf\_skip & 
t (s) & p & \\
\midrule

\textbf{Sequence of sanitization (Listing~\ref{lst:listing-1} at \textnormal{\circnum{7}}, \textnormal{\circnum{9}})}
& \textbf{1} & \textbf{SQLi \& PathTr.}
& \textbf{1.02 \& 1.01} & \textbf{1} & 3
& \textbf{-} & \textbf{0}
& \textbf{\toolFuzz{} detects SQLi as vulnerability and PathTr.\ as bug. Phuzz runs for 13 minutes} \\

\textbf{Safe Sequence (Listing~\ref{lst:listing-2} at \textnormal{\circnum{1}})}
& \textbf{2} & \textbf{Domain violation bug}
& \textbf{0.99} & \textbf{0} & 1
& \textbf{-} & \textbf{0}
& \textbf{Phuzz runs for 5 minutes} \\

Safe Sequence -- mysqli\_real\_escape\_string (Listing~\ref{lst:listing-2} at \circnum{2})
& 3 & --
& - & 0 & 1
& - & 0
& \toolFuzz{} safe-sequence check cancels function-based detection. Phuzz runs for 12 minutes \\

\textbf{Parameter in branch (Listing~\ref{lst:listing-2} at \textnormal{\circnum{4}})}
& \textbf{4} & \textbf{SQLi}
& \textbf{0.42} & \textbf{2} & 2
& \textbf{-} & \textbf{0}
& \textbf{Phuzz runs for 13 minutes} \\

\midrule

Arprice-responsive-pricing-table 3.6
& 5 & SQLi
& 282.42 & 20 & 914
& 176.98 & 1
& \\

Crm-perks-forms 1.0.7
& 6 & XSS
& 1.36 & 0 & 0
& 0.56 & 0
& \\

Essential-real-estate 3.9.5
& 7 & XSS
& 206.16 & 6 & 824
& 279.84 & 8
& \\

Gallery-album 1.9.9
& 8 & XSS
& 195.33 & 3 & 824
& 156.00 & 2
& \\

hypercomments 1.2.1
& 9 & XSS
& 15.02 & 2 & 787
& 21.71 & 2
& \\

\textbf{Joomsport-sports-league-results-management 5.1.7}
& \textbf{10} & \textbf{Deserialization}
& \textbf{176.05} & \textbf{3} & 824
& \textbf{-} & \textbf{3}
& \textbf{Phuzz cannot detect because input is base64-encoded. Phuzz runs for 11 minutes. \toolFuzz{} reports one False Positive for PathTr. due to file permission error.} \\

Kivicare-clinic-management-system 2.3.8
& 11 & SQLi
& 117.42 & 5 & 826
& 362.86 & 3
& \\

Nirweb-support 2.7.6
& 12 & SQLi
& 58.54 & 3 & 825
& 48.43 & 2
& Phuzz time is spent on syntax error (1064) \\

Nmedia-user-file-uploader 21.2
& 13 & PathTr.
& 88.50 & 0 & 810
& 16.97 & 1
& \\

Photo-gallery 1.6.2
& 14 & SQLi
& 120.50 & 5 & 824
& 303.10 & 7
& \\

Rezgo 4.1.6
& 15 & XSS
& 1.45 & 1 & 0
& 12.61 & 0
& \\

\textbf{Seo-local-rank 2.2.2}
& \textbf{16} & \textbf{PathTr.}
& \textbf{0.21} & \textbf{1} & 0
& \textbf{-} & \textbf{0}
& \textbf{Phuzz cannot detect due to \texttt{preg\_match} filtering. Phuzz runs for 14 minutes} \\

Show-all-comments-in-one-page 7.0.0
& 17 & XSS
& 209.09 & 7 & 824
& 171.52 & 3
& \\

\textbf{Totop-link 1.7}
& \textbf{18} & \textbf{Deserialization}
& \textbf{0.59} & \textbf{0} & 1
& \textbf{-} & \textbf{0}
& \textbf{Phuzz cannot detect due to base64-encoded input, runs for 12 minutes} \\

Ubigeo-peru 3.6.3
& 19 & SQLi
& 25.36 & 2 & 824
& 78.81 & 5
& Phuzz time is spent on syntax error (1064) \\

Udraw 3.3.2
& 20 & SQLi
& 5.28 & 2 & 824
& 12.36 & 2
& \\

Usc-e-shop 2.8.4
& 21 & SQLi
& 2.17 & 1 & 0
& 0.49 & 0
& \\

\midrule
Total & & & 1,267.63 & 66 & 8289 & 1,642.24 & 39 & \\ 
\bottomrule
\end{tabular}

\vspace{8pt} 
\noindent
\begin{minipage}{\linewidth} 
    \raggedright
    \small
    \hspace*{2.5em} \textbf{Note:}
    \begin{itemize}
        \item Total times are only for vulnerabilities or bugs detected by both \toolFuzz{} and Phuzz.
        \item A new path is counted when a vulnerability or bug is discovered, or, if none is found, when the fuzzer is stopped manually.
        \item For SQLi, Phuzz's detection time is recorded at the first syntax error, not at preceding errors that already indicate the presence of a bug.
    \end{itemize}
\end{minipage}

\end{table*}

\begin{table*}[t]
\centering
\caption{Wordpress Plugins Vulnerable Function Undetected by Phuzz}
\label{tab:wordpress_vuln_func}
\small
\begin{tabular}{lp{3cm}|p{13cm}}
\toprule
Test & Location & Vulnerable code \\
\midrule

10 &
\raggedright\path{joomsport-sports\-league-results\-management/includes/posts/joomsport-post-season.php:867} &
\begin{minipage}[t]{\linewidth}
\ttfamily
public static function joomsport\_md\_load() \{ \par
\quad \$mdId = intval(\$\_POST['mdId']); \par
\quad {\color{red}\textbf{\$args = unserialize(base64\_decode(\$\_POST['shattr']));}} //vuln function with encoded input \par
\quad \$args = array\_map('sanitize\_text\_field', \$args); \par
\}
\end{minipage}
\\
\midrule

16 &
\raggedright\path{seo-local-rank/admin/vendor/datatables/examples/resources/examples.php:3} &
\begin{minipage}[t]{\linewidth}
\ttfamily
<?php \par
{\color{blue}\textbf{if ( isset(\$\_POST['src']) \&\&}} \par
\quad {\color{blue}\textbf{preg\_match('/scripts\/[a-zA-Z\_\-]+\.php/', \$\_POST['src']) !== 0 ) \{ }}//guards vulnerability\par
//vuln function: file\_get\_contents \par
\quad {\color{red}\textbf{echo htmlspecialchars(file\_get\_contents('../server\_side/'.\$\_POST['src']));}} \par
\} else \{ \par
\quad echo ''; \par
\}
\end{minipage}
\\
\midrule

18 &
\raggedright\path{totop-link/totop-link.css.php:2} &
\begin{minipage}[t]{\linewidth}
\ttfamily
<?php \par
header('Content-type: text/css'); \par
{\color{red}\textbf{\$vars = unserialize(base64\_decode(\$\_GET['vars']));}} //vuln function with encoded input\par
\$width = (!empty(\$vars['width'])) ? \par
\quad 'width:'.\$vars['width'].'px;' : ''; \par
\end{minipage}
\\
\bottomrule

\end{tabular}
\vspace{2pt} 
    \begin{minipage}[t]{\textwidth}
        \small 
        Table~\ref{tab:wordpress_vuln_func} shows three example Wordpress plugin vulnerabilities (each from \texttt{/var/www/html/wordpress/wp-content/plugins/}) that Phuzz did not find but \toolFuzz{} found.
    \end{minipage}
\end{table*}


\begin{table*}[t]
\centering
\caption{Evaluation of \toolFuzz{} and Phuzz on DVWA; rce\_imposs is a non-vulnerability bug; p~=~new paths, t(s)= detection time, vf\_skip~=~the number of untainted vulnerable functions identified by the backward constant probe and skipped during function-level fuzzing.}
\label{tab:eval_dvwa}
\small
\begin{tabular}{lcl|rrr|rr|p{6.4cm}} 
\toprule
Benchmark &
Test &
Bug / &
\multicolumn{3}{c|}{\toolFuzz{}} & 
\multicolumn{2}{c|}{Phuzz} &
Info \\
&
No. &
Vuln. &
t (s) & p &  vf\_skip &
t (s) & p & \\
\midrule

sqli\_low & 22 & SQLi & 0.90 & 0 & 6 & 1.57 & 2 & \\
sqli\_med & 23 & SQLi & 0.91 & 0 & 7 & 1.44 & 2 & \\
sqli\_imposs & 24 & SQLi & - & 2 & 18 & - & 2 & \\
sqli\_blind\_low & 25 & SQLi & 1.50 & 0 & 6 & 1.50 & 2 & \\
sqli\_blind\_med & 26 & SQLi & 1.09 & 2 & 7 & 2.44 & 1 & \\
sqli\_blind\_high & 27 & SQLi & 1.80 & 4 & 6 & 31.17 & 5 & \\

rce\_low & 28 & RCE & 7.34 & 0 & 6 & 8.26 & 0 & \\
rce\_med & 29 & RCE & 1.68 & 0 & 6 & 1.75 & 0 & \\
rce\_high & 30 & RCE & 1.14 & 0 & 6 & 3.51 & 0 & \\
rce\_imposs & 31 & RCE (bug) & 7.68 & 1 & 12 & - & 1 & \toolFuzz{} uncovers bug \\

\textbf{fi\_low} & \textbf{32} & \textbf{PathTr.} & \textbf{0.81} & \textbf{0} & 6 & \textbf{-} & \textbf{0} &
\textbf{Phuzz cannot detect as its instrumentation cannot hook language construct such as include.} \\

\textbf{fi\_med} & \textbf{33} & \textbf{PathTr.} & \textbf{0.80} & \textbf{0} & 6 & \textbf{-} & \textbf{0} &
\textbf{Phuzz cannot detect as its instrumentation cannot hook language construct such as include.} \\

\textbf{fi\_high} & \textbf{34} & \textbf{PathTr.} & \textbf{1.48} & \textbf{2} & 3 & \textbf{-} & \textbf{1} &
\textbf{Phuzz cannot detect as its instrumentation cannot hook language construct such as include.} \\

fi\_imposs & 35 & PathTr. & - & - & 3 & - & 0 & \\

xss\_s\_low & 36 & XSS Stored & 1.54 & 0 & 7 & 6.87 & 0 & \\
xss\_s\_med & 37 & XSS Stored & 3.50 & 0 & 7 & 46.03 & 0 & \\
xss\_s\_high & 38 & XSS Stored & 11.00 & 0 & 7 & 9.70 & - & \\
xss\_s\_imposs & 39 & XSS Stored & - & - & 7 & - & 0 & \\

xss\_r\_low & 40 & XSS Reflected & 1.87 & 1 & 7 & 2.81 & 0 & \\
xss\_r\_med & 41 & XSS Reflected & 2.04 & 1 & 7 & 2.92 & 0 & \\
xss\_r\_high & 42 & XSS Reflected & 1.35 & 2 & 7 & 1.00 & 0 & \\
xss\_r\_imposs & 43 & XSS Reflected & - & 2 & 7 & - & 0 & \\

\midrule

Total & & & 37.66 & 17 & 159 & 120.97 & 16 & \\ 
\bottomrule
\end{tabular}
\vspace{8pt} 
\begin{minipage}{\linewidth} 
    \raggedright
    \small
    \hspace*{2.5em} \textbf{Note:}
    \begin{itemize}
        \item Total times are only for vulnerabilities or bugs detected by both \toolFuzz{} and Phuzz.
        \item A new path is counted when a vulnerability or bug is discovered, or, if none is found, when the fuzzer is stopped manually.
        \item For SQLi, Phuzz's detection time is recorded at the first syntax error, not at preceding errors that already indicate the presence of a bug.
        \item DVWA sqli\_high is not included because Phuzz and \toolFuzz{} do not support second order vulnerability detection currently.
    \end{itemize}
\end{minipage}
\end{table*}

\begin{table*}[t]
\centering
\caption{Evaluation of \toolFuzz{} and Phuzz on bWAPP; p~=~new paths, t(s)= detection time, vf\_skip~=~the number of untainted vulnerable functions identified by the backward constant probe and skipped during function-level fuzzing.
}
\label{tab:eval_bwapp}
\small
\begin{tabular}{lcl|rrr|rr|p{6cm}} 
\toprule
Benchmark &
Test &
Bug / &
\multicolumn{3}{c|}{\textbf{\toolFuzz{}}} &
\multicolumn{2}{c|}{\textbf{Phuzz}} &
Info \\
&
&
Vuln.&
t (s) & p & vf\_skip &
t (s) & p & \\
\midrule

commandi & 44 & RCE & 2.00 & 1 & 10 & 2.61 & 0 & \\
commandi\_blind & 45 & RCE & 0.90 & 1 & 10 & 4.17 & 0 & \\

\textbf{pathtraversal} & \textbf{46} & \textbf{PathTr.} & \textbf{0.96} & \textbf{1} & 9 & \textbf{-} & \textbf{0} &
\textbf{Phuzz has an FP as it triggers error at \texttt{is\_file}, which is a filtering function, not a vulnerable one to allow only valid file path; \toolFuzz{} detects via function-trace-based check because no error produced.} \\

\textbf{phpi} & \textbf{47} & \textbf{RCE} & \textbf{0.64} & \textbf{1} & 7 & \textbf{-} & \textbf{1} &
\textbf{Phuzz cannot detect as its instrumentation cannot hook language construct such as eval.} \\

\textbf{rlfi} & \textbf{48} & \textbf{PathTr.} & \textbf{0.63} & \textbf{0} & 10 & \textbf{-} & \textbf{0} &
\textbf{Phuzz cannot detect as its instrumentation cannot hook language construct such as include.} \\

sqli\_ajax & 49 & SQLi & 0.52 & 0 & 9 & 5.49 & 1 & \\
sqli\_blind\_boolean & 50 & SQLi & 0.18 & 0 & 13 & 16.97 & 1 & \\
sqli\_blind\_time & 51 & SQLi & 0.74 & 1 & 13 & 5.35 & 1 & \\
sqli\_get\_search & 52 & SQLi & 0.56 & 1 & 13 & 6.91 & 1 & \\
sqli\_get\_select & 53 & SQLi & 1.36 & 2 & 14 & 0.83 & 0 & \\
sqli\_login\_hero & 54 & SQLi & 0.53 & 0 & 13 & 6.07 & 2 & \\
sqli\_post\_search & 55 & SQLi & 0.36 & 1 & 13 & 19.43 & 1 & \\
sqli\_post\_select & 56 & SQLi & 0.89 & 2 & 12 & 0.71 & 1 & \\
sqli\_stored\_blog & 57 & SQLi & 0.11 & 3 & 17 & 1.06 & 0 & \\

xss\_ajax & 58 & XSS & 0.56 & 1 & 5 & 0.92 & 0 & \\
xss\_back\_button & 59 & XSS & 0.23 & 0 & 10 & 12.51 & 0 & \\
xss\_custom\_header & 60 & XSS & 1.02 & 0 & 10 & 5.80 & 0 & \\
xss\_eval & 61 & XSS & 0.74 & 0 & 9 & 8.81 & 0 & \\
xss\_get & 62 & XSS & 0.22 & 1 & 10 & 4.24 & 0 & \\
xss\_href & 63 & XSS & 1.42 & 1 & 14 & 4.19 & 0 & \\
xss\_json & 64 & XSS & 0.72 & 1 & 10 & 0.38 & 0 & \\
xss\_post & 65 & XSS & 0.50 & 1 & 10 & 21.53 & 0 & \\
xss\_referer & 66 & XSS & 0.18 & 0 & 10 & 7.57 & 0 & \\
xss\_stored & 67 & XSS & 1.32 & 1 & 17 & 0.65 & 0 & \\
xss\_useragent & 68 & XSS & 0.60 & 0 & 10 & 12.48 & 0 & \\
\midrule
Total & & & 15.66 & 20 & 278 & 148.68 & 9 & \\ 
\bottomrule
\end{tabular}
\vspace{8pt} 
\begin{minipage}{\linewidth} 
    \raggedright
    \small
    \hspace*{2.5em} \textbf{Note:}
    \begin{itemize}
        \item Total times are only for vulnerabilities or bugs detected by both \toolFuzz{} and Phuzz.
        \item A new path is counted when a vulnerability or bug is discovered, or, if none is found, when the fuzzer is stopped manually.
        \item For SQLi, Phuzz's detection time is recorded at the first syntax error, not at preceding errors that already indicate the presence of a bug.
    \end{itemize}
\end{minipage}
\end{table*}

\begin{table*}[t]
\centering
\caption{bWAPP Vulnerable Function Undetected by Phuzz}
\label{tab:bwapp_vuln_func}
\small
\begin{tabular}{l|p{2cm}|p{14cm}}
\toprule
Test & Location & Vulnerable code \\
\midrule
46 &
\raggedright\path{/var/www/html/bwapp/directory_traversal_1.php:141} &
\begin{minipage}[t]{\linewidth}
\ttfamily
function show\_file(\$file) \{ \par
\quad // Checks whether a file or directory exists \par
\quad {\color{blue}\textbf{if(is\_file(\$file))}} //blockage, only allows valid path, leading to no error\par
\quad \{ //fopen is vuln, detectable only function-trace-based check\par
\quad\quad {\color{red}\textbf{\$fp = fopen(\$file, "r") or die("Couldn't open \$file.");}} \par
\quad\quad while(!feof(\$fp)) \{ \par
\end{minipage}
\\

\hline
\end{tabular}
\end{table*}

\begin{table*}[t]
\caption{Evaluation of \toolFuzz{} and Phuzz on wackopicko (top) and XVWA (bottom); p~=~new paths, t(s)= detection time, vf\_skip~=~the number of untainted vulnerable functions identified by the backward constant probe and skipped during function-level fuzzing.}
\label{tab:eval_wacko_xvwa}
\centering
\small
\begin{tabular}{p{3cm}cl|rrr|rr|p{5cm}} 
\toprule
Benchmark &
Test &
Bug/ &
\multicolumn{3}{c|}{\toolFuzz{}} &
\multicolumn{2}{c|}{Phuzz} &
Info \\
&
&
Vulnerability &
t (s) & p & vf\_skip &
t (s) & p & \\
\midrule

\textbf{admin}
& \textbf{69} & \textbf{PathTr.}
& \textbf{0.27} & \textbf{0} & 0
& \textbf{--} & \textbf{0}
& \textbf{Phuzz cannot detect as its instrumentation cannot hook language construct such as require\_once.} \\

guestbook
& 70 & XSS
& 14.52 & 1 & 8
& 1.27 & 0
&  \\

login
& 71 & SQLi
& 0.15 & 1 & 16 
& 24.22 & 2
&  \\

passcheck
& 72 & RCE
& 0.09 & 1 & 6
& 14.22 & 1
&  \\

piccheck
& 73 & XSS
& 0.93 & 0 & 7
& 11.96 & 0
&  \\

\textbf{register\_1 (\$vuln set to True in codebase)}
& \textbf{74} & \textbf{SQLi}
& \textbf{1.48} & \textbf{3} & 8
& \textbf{--} & \textbf{0}
& \textbf{Phuzz cannot detect due to inability to satisfy branch condition containing input parameter.} \\

search
& 75 & XSS
& 3.49 & 2 & 13
& 28.16 & 2
&  \\

submitname
& 76 & XSS
& 0.40 & 0 & 7
& 13.61 & 0
&  \\

\textbf{upload\_1 (added, not in Phuzz existing input)}
& \textbf{77} & \textbf{PathTr.}
& \textbf{16.98} & \textbf{8} & 11
& \textbf{--} & \textbf{0}
& \textbf{Phuzz cannot detect due to inability to generate file/multi-form-data content} \\

\midrule

\textbf{fi}
& \textbf{78} & \textbf{PathTr.}
& \textbf{0.26} & \textbf{0} & 6
& \textbf{--} & \textbf{0}
& \textbf{Phuzz cannot detect as its instrumentation cannot hook language construct such as include.} \\

phpobject
& 79 & Deserialization
& 0.63 & 0 & 6
& 0.78 & 0
&  \\

rce
& 80 & RCE
& 0.70 & 1 & 6
& 2.68 & 0
&  \\

sqli\_blind\_item
& 81 & SQLi
& 0.67 & 2 & 6
& 0.53 & 0
&  \\

sqli\_blind\_search
& 82 & SQLi
& 0.23 & 2 & 8
& 4.21 & 0
&  \\

sqli\_item
& 83 & SQLi
& 0.72 & 3 & 6
& 0.12 & 0
&  \\

sqli\_search
& 84 & SQLi
& 0.44 & 2 & 8
& 1.50 & 1
&  \\

xss\_stored
& 85 & XSS Stored
& 1.20 & 3 & 7
& 0.69 & 0
&  \\

xss\_reflected
& 86 & XSS Reflected
& 0.67 & 0 & 6
& 7.51 & 0
&  \\

\midrule
Total & & & 24.84 & 29 & 135 & 111.46 & 6 & \\ 
\bottomrule
\end{tabular}
\vspace{8pt} 
\begin{minipage}{\linewidth} 
    \raggedright
    \small
    \hspace*{2.5em} \textbf{Note:}
    \begin{itemize}
        \item Total times are only for vulnerabilities or bugs detected by both \toolFuzz{} and Phuzz.
        \item A new path is counted when a vulnerability or bug is discovered, or, if none is found, when the fuzzer is stopped manually.
        \item For SQLi, Phuzz's detection time is recorded at the first syntax error, not at preceding errors that already indicate the presence of a bug.
    \end{itemize}
\end{minipage}

\end{table*}

\begin{table*}[t]
\centering
\caption{wackopicko Vulnerable Function Undetected by Phuzz}
\label{tab:wackopicko_vuln_func}
\small
\begin{tabular}{l|p{1.5cm}|p{14.5cm}}
\hline
\textbf{Test} & \textbf{Location} & \textbf{Vulnerable code} \\
\hline

74 &
Entry point: \par
\raggedright\path{/var/www/html/wackopicko/website/users/register.php:19} &
\begin{minipage}[t]{\linewidth}
\ttfamily
//input parameters: password and againpass are in branch \par
\textcolor{blue}{if (\$\_POST['password'] != \$\_POST['againpass']) \{} // vulnerability triggered if password == againpass \par
\quad \$flash['error'] = "The passwords do not match. Try again"; \par
\quad \$error = True; \par
\} \par
// vulnerable call is triggered when flag is True \par
\textcolor{blue}{else if (\$new\_id = Users::create\_user(\$\_POST['username'], \$\_POST['password'],} \par
\quad \$\_POST['firstname'], \$\_POST['lastname'], True)) \{ \par
\quad Users::login\_user(\$new\_id); \par
\quad http\_redirect(Users::\$HOME\_URL); \par
\}
\end{minipage}
\\
\hline

74 &
Vulnerable function: \par
\raggedright\path{/var/www/html/wackopicko/website/include/users.php:53} &
\begin{minipage}[t]{\linewidth}
\ttfamily
static function create\_user(\$username, \$pass, \$firstname, \$lastname, \$vuln = False) \{ \par
\quad \$salt = mt\_rand(0, 900); \par
\quad \$salt = base64\_encode(\$salt); \par
\quad if (\$vuln) \{ \par
\quad\quad \$pass = mysqli\_real\_escape\_string(OURDB->conn, \$pass); \par
\quad\quad \$firstname = mysqli\_real\_escape\_string(OURDB->conn, \$firstname); \par
\quad\quad \$pass = \$pass . \$salt; \par
\quad\quad \$query = "INSERT INTO users ... VALUES (..., '\{\$username\}', SHA1('\{\$pass\}'), '\{\$firstname\}', ... );" \par
\quad \} \par
\}
// other codes \par
\quad \textcolor{red}{(\$res = mysqli\_query(OURDB-\textgreater conn, \$query)) // vulnerable function} \par
\end{minipage}
\\

\hline
\end{tabular}
\end{table*}

\begin{table*}[t]
\vspace{0.75cm}
\caption{Write Volume Details, Time, and Write Time of \toolInstr{} and PCOV + UOPZ for the first HTTP request of the original (unmutated) input averaged over 10 samples;
E+E~=~Errors \& exceptions; Trace~=~vulnerable function and construct trace
Aprice~=~Arprice-responsive-pricing-table 3.6 WordPress plugin.
}
\label{tab:write-details-time}
\centering
\small
\resizebox{\textwidth}{!}{%
\begin{tabular}{l|rrrrrrr|rrrr}
\toprule
Benchmark &
\multicolumn{7}{c|}{\toolInstr{}} &
\multicolumn{4}{c}{PCOV + UOPZ} \\

& Coverage & E+E & Trace & Tainted- & Total & Time & Write 
& Coverage & E+E & Total & Time  \\
& (kB) & (kB) & (kB) & branches(kB) & (kB) & (ms) & (ms) 
& (kB) & (kB) & (kB) & (ms)  \\
\midrule

\texttt{poc\_seq\_sanit} 
& 0.10 & -- & 1.30 & -- & 1.401 & 6.2 & 1.8
& 0.35 & 0.06 & 0.407 & 6.0  \\

\texttt{Arprice} 
& 6,656.00 & 13.70 & 22,804.8 & 2.10 & 29,476.280 & 4,166.9 & 657.4 
& 5,427.20 & 375.73 & 5,802.928 & 6,553,5 \\

\texttt{dvwa\_sqli\_low} 
& 1.10 & -- & 3.50 & -- & 4.600 & 4.2 & 0.3 
& 16.30 & 1.61 & 17.913 & 678.5  \\

\texttt{bwapp\_sqli\_ajax} 
& 0.25 & -- & 1.90 & -- & 2.153 & 11.0 & 0.4 
& 2.50 & 0.39 & 2.888 & 12.9  \\

\texttt{wacko\_register\_1} 
& 0.44 & -- & 2.40 & 1.40 & 4.242 & 8.4 & 0.6 
& 1.80 & 0.50 & 2.302 & 10.7  \\

\texttt{xvwa\_sqli\_search} 
& 0.16 & -- & 1.90 & 0.25 & 2.309 & 12.1 & 0.5 
& 0.93 & 0.40 & 1.335 & 14.7  \\

\texttt{instrumentation test-1-a \circnum{1}} 
& 0.05 & -- & -- & -- & 0.055 & 0.3 & 0.1 
& 0.84 & -- & 0.844 & 1.1  \\

\texttt{instrumentation test-1-b \circnum{2}} 
& 0.05 & -- & -- & -- & 0.055 & 0.3 & 0.1 
& 0.03 & -- & 0.033 & 0.8  \\

\texttt{instrumentation test-2 \circnum{3}} 
& 0.15 & -- & -- & -- & 0.146 & 0.5 & 0.1 
& 0.15 & -- & 0.152 & 1.4  \\

\bottomrule
\end{tabular}%
}
\noindent
\begin{minipage}{\linewidth} 
    \raggedright
    \small
    \hspace*{2.5em} \textbf{Note:}: The last three inputs aim to illustrate the efficiency of \toolInstr{} over PCOV plus UOPZ using inputs with single execution path without any branch instructions and userlands functions to be monitored, where \toolInstr{} performs much faster than them.
    \begin{itemize}
        \item \circnum{1} Target program is given by Listing~\ref{lst:instr-test-1} below. For this test, PCOV and UOPZ are activated. PCOV and UOPZ global hooks are invoked. On the other hand, \toolInstr{} hooks are never invoked; the coverage data is only for the file path of the triples; see Section \ref{coverage-triples} for the definition of execution path as (file path, branch opcode’s userland line number, branch-opcode-outcome) triples.
        \item \circnum{2} Target program is given by Listing~\ref{lst:instr-test-1} below. For this test, PCOV is not activated and UOPZ is activated. This measures the overhead caused by UOPZ globally intercepting Zend’s call dispatchers that takes effect even when only invulnerable function exits in web application, such as \texttt{add} function that is never hooked. \toolInstr{} hooks are never invoked because no branch instructions in the code and \toolInstr{} does not hook Zend's call dispatcher to monitor function, but directly hooks the handler of vulnerable and sanitization functions. 
        \item \circnum{3} Target program is given by Listing~\ref{lst:instr-test-2} below. This measures the impact of looping for \toolInstr{} that hooks the conditional statement in the loop header only, while PCOV global hook is called for every instruction and UOPZ hook is invoked for Zend's call dispatcher.
    \end{itemize}
\end{minipage}
\end{table*}

\clearpage
\begin{figure*}[!t]
\centering
\begin{minipage}[t]{0.49\textwidth}
\begin{lstlisting}[language=PHP,
basicstyle=\small\ttfamily,
caption={Instrumentation test-1},
label={lst:instr-test-1},
numbers=none,
captionpos=b]
<?php
/*To show the overhead of PCOV performing global hooking of opcode despite only one path executed and UOPZ performing global hooking of call dispatchers despite add, an invulnerable function, is never hooked. \toolInstr{} hooks are never invoked since there is no branches and no vulnerable/sanitization functions.
*/ 
$total = 0;
$i = 1;

//add is called 100 times
add($i);
add($i);
add($i);
/*
 *
 *other calls
 *
 *
*/
add($i);

function add($number) {
    global $total;
    $total += $number;
}
?>
\end{lstlisting}
\end{minipage}
\hfill
\begin{minipage}[t]{0.49\textwidth}
\begin{lstlisting}[language=PHP,numbers=left, xleftmargin=2em, xrightmargin=1em,
label={lst:instr-test-2},
basicstyle=\small\ttfamily,
caption={Instrumentation test-2},
numbers=none,
captionpos=b]
<?php
/*To illustrate the time spent by \toolInstr{} for logging the branch statement in loop header, and PCOV time for logging every executed instruction.
*/
//maxcounter input is set to 100 to match Listing 3}
$maxcounter = isset($_GET['maxcounter']) ? (int)$_GET['maxcounter'] : 0;
$total = 0;

for ($i = 0; $i < 3; $i++) {
    add(1);
}

function add($number) {
    global $total;
    $total += $number;
}
?>
\end{lstlisting}
\end{minipage}
\end{figure*}

\begin{figure*}[h!]
    \vspace*{0.75cm}
    \centering
    \includegraphics[width=\textwidth]{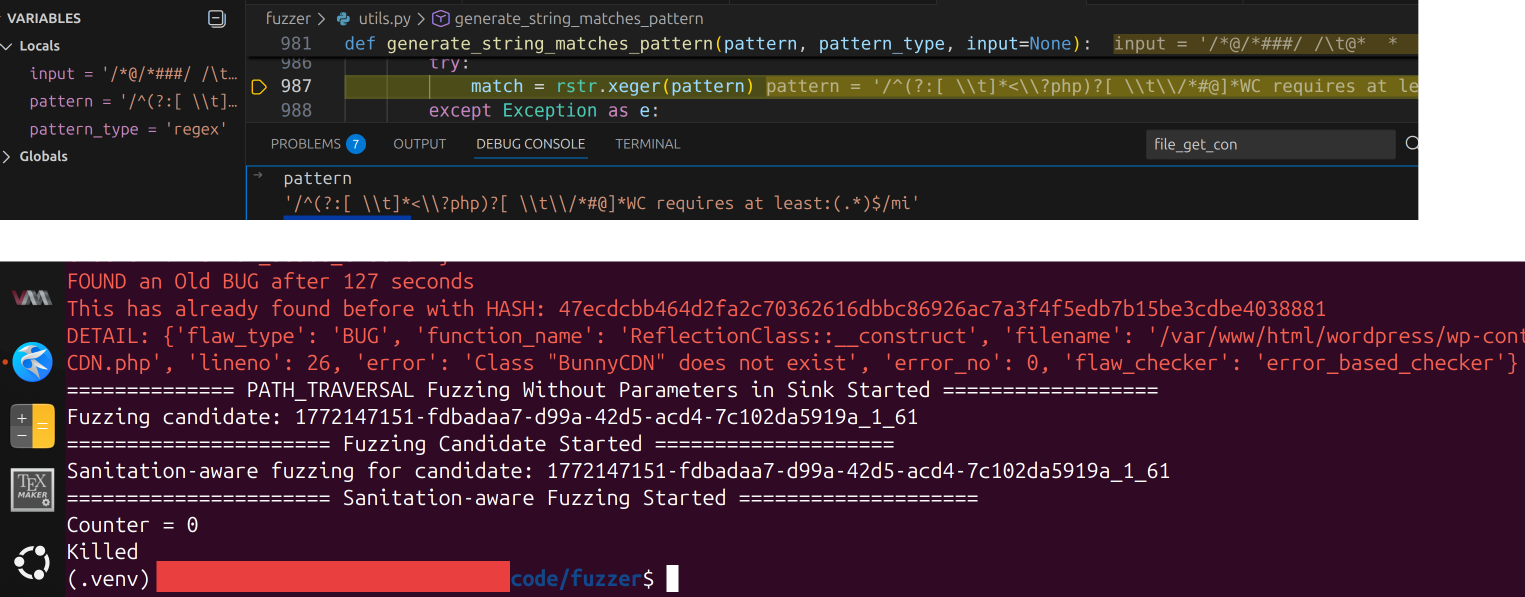}
    \caption{Failure of \texttt{xeger} on complex regex patterns leading to Python termination.}
    \label{fig:xeger_failure}

    \vspace{2pt}
    \begin{minipage}{\textwidth}
        \textit{Note:} pattern used in experiment:
            \begin{verbatim}
            /^(?:[ \t]*<\?php)?[ \t\/*#@]*WC requires at least:(.*)$/mi
            \end{verbatim}
    \end{minipage}
\end{figure*}

\end{document}